\documentclass[twocolumn]{aastex63}
\newcommand{\ha}{H$\alpha$}
\newcommand{\HI}{\ion{H}{1}~}

\newcommand{\hii}{H\,{\small II}}

\newcommand{\kms}{km~s$^{-1}$}

\newcommand{\Lya}{Ly$\alpha$~}

\newcommand{\mhi}{$\Sigma_{\rm{H\,{\small I}}}$}

\newcommand{\mstar}{M$_\star$}
\newcommand{\p}{$^\prime$}

\newcommand{\sdssr}{{\em r}}
\newcommand{\sdssg}{{\em g}}
\newcommand{\sfr}{$\Sigma_{\rm{SFR}}$}
\newcommand{\sm}{$\Sigma_{\star}$}

\newcommand{\lamlam}{$\rm \lambda\lambda$}

\shorttitle{The DIISC Survey Overview}
\shortauthors{Borthakur et al.}

\begin{document}

\title{DIISC Survey: Deciphering the Interplay Between the Interstellar Medium, Stars, and the Circumgalactic Medium Survey}

%\correspondingauthor{Sanchayeeta Borthakur}
%\email{sanchayeeta.borthakur@asu.edu}
\author[0000-0002-2724-8298]{Sanchayeeta Borthakur}
\affiliation{School of Earth and Space Exploration, Arizona State University, 781 Terrace Mall, Tempe, AZ 85287, USA}

\author[0000-0002-3472-0490]{Mansi Padave}
\affiliation{School of Earth and Space Exploration, Arizona State University, 781 Terrace Mall, Tempe, AZ 85287, USA}

\author[0000-0001-6670-6370]{Timothy Heckman}
\affiliation{Center for Astrophysical Sciences, Department of Physics and Astronomy, Johns Hopkins University, Baltimore, MD 21218, USA}
\affiliation{School of Earth and Space Exploration, Arizona State University, 781 Terrace Mall, Tempe, AZ 85287, USA}
 
\author[0000-0003-1436-7658]{Hansung B. Gim}
\affiliation{Department of Physics, Montana State University, P. O. Box 173840, Bozeman, MT 59717, USA}
\affiliation{School of Earth and Space Exploration, Arizona State University, 781 Terrace Mall, Tempe, AZ 85287, USA}
 
\author[0000-0002-2819-0753]{Alejandro J. Olvera}
\affiliation{School of Earth and Space Exploration, Arizona State University, 781 Terrace Mall, Tempe, AZ 85287, USA}

\author[0000-0001-5530-2872]{Brad Koplitz}
\affiliation{School of Earth and Space Exploration, Arizona State University, 781 Terrace Mall, Tempe, AZ 85287, USA}

\author[0000-0003-3168-5922]{Emmanuel Momjian} 
\affiliation{National Radio Astronomy Observatory, 1003 Lopezville Road, Socorro, NM 87801, USA}

\author[0000-0003-1268-5230]{Rolf A. Jansen}
\affiliation{School of Earth and Space Exploration, Arizona State University, 781 Terrace Mall, Tempe, AZ 85287, USA}

\author[0000-0002-8528-7340]{David Thilker} 
\affiliation{Center for Astrophysical Sciences, Department of Physics and Astronomy, Johns Hopkins University, Baltimore, MD 21218, USA}

\author{Guinevere Kauffman} 
\affiliation{Max-Planck Institut f\"{u}r Astrophysik, D-85741 Garching, Germany}

\author[0000-0003-0724-4115]{Andrew J. Fox} 
\affiliation{AURA for ESA, Space Telescope Science Institute, 3700 San Martin Drive, Baltimore, MD 21218, USA}
\affiliation{Department of Physics \& Astronomy, Johns Hopkins University, 3400 N. Charles Street, Baltimore, MD 21218, USA}

\author[0000-0002-7982-412X]{Jason Tumlinson} 
\affiliation{Space Telescope Science Institute, 3700 San Martin Drive, Baltimore, MD 21218, USA}

\author[0000-0001-5448-1821]{Robert C.Kennicutt} 
\affiliation{Department of Astronomy and Steward Observatory, University of Arizona Tucson, AZ 85721, USA}
\affiliation{Department of Physics and Astronomy, Texas A\&M University College Station, TX 77843, USA}

\author[0000-0001-8421-5890]{Dylan Nelson} 
\affiliation{Universit\"{a}t Heidelberg, Zentrum f\"{u}r Astronomie, Institut f\"{u}r theoretische Astrophysik, Albert-Ueberle-Str. 2, D-69120 Heidelberg, Germany}

\author[0000-0002-2724-8298]{Jacqueline Monckiewicz}
\affiliation{School of Earth and Space Exploration, Arizona State University, 781 Terrace Mall, Tempe, AZ 85287, USA}

\author[0000-0002-7314-2558]{Thorsten Naab} 
\affiliation{Max-Planck Institut f\"{u}r Astrophysik, D-85741 Garching, Germany}

\begin{abstract}

 We present the Deciphering the Interplay between the Interstellar medium, Stars, and the Circumgalactic medium (DIISC) Survey. This survey is designed to investigate the correlations in properties between the circumgalactic medium (CGM), the interstellar medium (ISM), stellar distributions, and young star-forming regions. The galaxies were chosen to have a QSO sightline within 3.5 times the \HI radii probing the disk-CGM interface.
The sample contains 34 low-redshift galaxies with a median stellar mass of 10$^{10.45}~\rm M_{\odot}$ probed at a median impact parameter of $\rho=$55~kpc. The survey combines ultraviolet spectroscopic data from the Cosmic Origins Spectrograph aboard the Hubble Space Telescope with \HI\ 21~cm hyperfine transition imaging with the Very Large Array (VLA), ultraviolet imaging with Galaxy Evolution Explorer (GALEX), and optical imaging and spectroscopy with the MMT and Vatican Advanced Technology Telescope. We describe the specific goals of the survey, data reduction, high-level data products, and some early results.
We present the discovery of a strong inverse correlation, at a confidence level of 99.99\%, between \Lya equivalent width, $\rm W_{Ly\alpha}$, and impact parameter normalized by the \HI radius ($\rho/R_{HI}$).
We find $\rho/R_{HI}$ to be a better empirical predictor of \Lya equivalent width than virial radius normalized impact parameter ($\rho/R_{vir}$) or parameterizations combining $\rho,~R_{vir}$, stellar mass, and star formation rate. We conclude that the strong anticorrelation between the \Lya equivalent width and $\rho/R_{HI}$ indicates that the neutral gas distribution of the CGM is more closely connected to the galaxy's gas disk rather than its stellar and dark matter content.

\end{abstract}


\section{Introduction} \label{sec:intro}

Throughout cosmic time, galaxy disks contain a significant fraction of cold gas despite continuously converting gas into stars \citep{Carilli_walter13}. This is surprising as the gas depletion times ($\rm M_{gas}/SFR$) for most high-redshift galaxies are close to a Gyr \citep[][and references therein]{Tacconi18}. Yet, galaxies at present times contain large reservoirs of neutral interstellar medium (ISM) in their disks \citep{Zwaan97, Jones18}. These observations require that galaxy disks acquire gas throughout cosmic time to counter the loss due to star formation and maintain their neutral gas disks. Understanding the processes of gas flows that feed and deplete gas disks of galaxies is crucial in understanding how galaxies evolve. 
The study of these processes was also featured as one of the main recommendations (Cosmic Ecosystems) of the Decadal Survey on Astronomy and Astrophysics 2020 \citep{Astro2020} as a focus area this decade.

One of the foremost gas and metal reservoirs that support gas accretion into galaxies is the circumgalactic medium \citep[CGM;][]{Tumlinson17, Peroux_howk20}. 
The CGM contains as much mass as the stellar mass within the galaxy \citep{werk14} and occupies 99\% of the volume of the dark matter halo.
The CGM is the only path that connects the disk to the intergalactic medium (IGM).
Therefore, any gas flow to and from the disks has to move through the CGM.
It is not surprising that the CGM's neutral gas content strongly correlates with the \HI 21cm mass tracing the interstellar medium within the disks \citep{borthakur15}. While this correlation likely indicates gas accretion, the physics of how gas is transported from the CGM and incorporated into the disks has still eluded us.

The disk-CGM interface holds a crucial place for studies of gas cycling in and out of the disk. The gas disk, which is often larger than the stellar disk of star-forming galaxies \citep{bosma17}, tends to drop in column density rapidly as it hits column density of $\rm 10^{19} cm^{-2}$ \citep{Corbelli93,Ianjamasimanana18, Sardone21}, thus making it difficult to detect with our current radio interferometers. In addition to the extended disk, a significant fraction ($\sim \rm 14\%$) total \HI\ of disk galaxies exists as extraplanar gas that extends above and below the disk \citep[e.g., HALOGAS survey;][]{heald11, Marasco19}. Some of these clouds show offsets from the co-rotating disk and have been referred to as anomalous velocity clouds \citep[e.g.,][]{thilker04, gim21} or as high- and intermediate velocity clouds (HVCs) in the Milky Way and nearby galaxies \citep{wakker97, hartmann97,kalberla08, putman12}.
Absorption line spectroscopic studies of the Milky Way HVCs using background Quasi Stellar Object (QSO) show that HVCs are bringing $\rm M_{total}^{acc} \approx 0.8-1.4~M_{\odot}~yr^{-1}$ \citep{lehner11} towards the disk.

It is still debated how the inflowing gas gets incorporated into the disk \citep{somerville_dave15, Crain_vandeVoort23}. 
One set of theoretical studies broadly finds that gas accretes onto the extended disk and gradually moves radially inward \citep[e.g.][]{Stewart11, hafen22, trapp22, stern24}. 
 Once the circumgalactic gas settles into the outer disk, it moves radially inwards to facilitate star formation.
Theoretical models suggest that such flows would be mildly subsonic with velocities of a few \kms~ in the inner disks \citep{nuza19, okalidis21, trapp22, trapp24}. 
However, the radial inflow velocities in the outer disk are expected to be much higher to keep the same mass flux, making the extended disk and disk-CGM interface perfect laboratories for detecting radial flows.
\citet{schmidt16} discovered radial gas flows in a few nearby galaxies, e.g., 15~\kms\ flow carrying a mass flux of 3$ \rm~M_{\odot}~yr^{-1}$ in NGC~2403. 
Conversely, \citet{di_teodoro_peek22} found an average mass flow rate of just 0.3$ \rm~M_{\odot}~yr^{-1}$ for a larger sample.

The second set of theoretical studies suggests that a significant fraction of gas accreted in present-day Milky Way-type galaxies is via recycled accretion. Material ejected from the disk in past star-formation feedback may return to the disk to fuel future star formation \citep{Oppenheimer08, ford14, Angles-Alcazar17, vandeVoort17, Grand19}. A proposed framework, often termed fountain flow, theorizes that hot metal-rich outflows interact with the CGM, which enables gas cooling and eventually leads to the return of these cooler clouds towards the disk \citep{Shapiro76, Bregman80, Fraternali17, gas_accretion17}.  For example, \citet{Fraternali15} showed that the kinematics of Complex C in the Milky Way halo is consistent with it being expelled from the Cygnus-Outer spiral arm about 150 Myr ago and is now returning towards the disk after triggering condensation of the warm CGM onto it.

While both modes of gas accretion are consistent with some observations, there are still crucial inconsistencies. \citet{Afruni21, Afruni23} found that the cool clouds observed in the CGM of low-redshift galaxies \citep[e.g.][and by others]{chen98, werk14, borthakur16, Keeney17, lehner19, Lehner20, chen20, wilde21, Nateghi23a, Klimenko23}
can not be reproduced entirely via fountain flows generated by supernova feedback nor by accretion from the IGM alone.  
It is not surprising, as the accretion rate is expected to be delicately balanced between the force of gravitation and forces generated by star formation and AGN-driven feedback. Magnetic fields may also play a role \citep{Heesen23, Pakmor20, vandeVoort21}; we have yet to nail down all the observable, although recent studies show promising results \citep{Buie22, Casavecchia24}.

Multiple studies have attempted to characterize CGM properties in the context of star formation within their host galaxies in the last decade.
The COS-Halos survey found that the amount of highly ionized gas in the CGM traced by the \ion{O}{6} transition correlates with the specific star formation rate (sSFR) of the galaxies \citep{Tumlinson11}. 
\citet{werk16} suggested that the broad \ion{O}{6} absorbers seen in the COS-Halos sample are either tracing highly structured clouds that are primarily photoionized by local high energy sources (as opposed to the cosmic UV background) or gas radiatively cooling behind supersonic winds. 
Later, theoretical studies also suggested that the correlation of \ion{O}{6} and sSFR reflects the halo mass trend of the sample, likely modulated by the impact of AGN feedback \citep{oppenheimer16, nelson18}.
Other observational studies have also found correlations between other ionized species in the CGM and the specific star formation rate of the galaxies \citep{borthakur13, heckman17, Garza24}. However, a study of the CGM properties of AGN-dominated galaxies by \citep{berg18} showed a supressed detection rate of all species. This led the authors to infer that either the AGN host galaxies have a fundamentally different CGM than non-AGN dominated galaxies or there had been past starburst activity in those galaxies that may have triggered the AGN.

In summary, there is ample evidence that multiple processes can be concurrently active in galaxies and impact the CGM, especially at the disk-CGM interface, where complex gas physics balances the rates of accretion and outflow and likely modulates star formation and AGN activity.
Even for a green-valley galaxy like the Milky Way, there is evidence of extended disks \citep{kalberla09}, star-formation and AGN-driven feedback \citep{fox15}, galaxy interaction impacting the CGM \citep{smart23}, and gas inflows \citep{putman11, putman12}.
Therefore, we expect that the gas flows between the CGM, the ISM, and stars are supported by multi-phase baryonic processes common in most galaxies. However, our knowledge of these processes in other galaxies is limited. Most 21cm \HI studies focus on nearby galaxies, and most low-redshift CGM surveys cover galaxies beyond the redshift range of 21cm surveys \citep{giovanelli05}. Even for low-redshift CGM studies centered around $z=0.1$, the galaxies are not well resolved (at the scales of individual clusters or stellar associations) to correlate CGM properties with spatially resolved star formation or have the sensitivity look for lower mass satellite galaxies around them \citep[e.g.][]{Geha17}.

To get the complete picture of how the CGM, the ISM, and stars are connected in the baryon cycle, we must study the processes occurring in these phases simultaneously. The first step in this direction would be to have a sample of galaxies where we have access to the resolved maps of the ISM, stars, and star-forming activity that can be correlated with the properties of the CGM. So far, we have not done that in a statistically significant manner. In the era of integral field units (IFU), CGM studies of the galaxies from the MANGA survey \citep{Klimenko23} and other IFU and long-slit follow-up of galaxies from CGM studies offer a powerful way to connect the stellar disk with the CGM. However, even with IFU, most galaxies can not be resolved to sub-kpc scales, nor do we get a complete census of the cool ISM fueling star formation. 
To fill this gap and trace the complete baryon cycle in and around the disk of galaxies, we have designed the Deciphering the Interplay between the Interstellar medium, Stars, and the Circumgalactic medium (DIISC) Survey.

The DIISC survey consists of three main components that trace the CGM, the ISM, and stars in a sample of 33 low-redshift galaxies. The main objectives of the survey are to identify correlations between the properties of the CGM, the ISM, and stars and relate that to the flow of gas in the context of the baryon cycle. To enable these objectives, we have acquired QSO absorption line spectroscopy with the Cosmic Origins Spectrograph (COS) aboard the Hubble Space Telescope (termed the COS-DIISC program), 21~cm HI interferometric data with the Karl G. Jansky Very Large Array \footnote{The National Radio Astronomy Observatory is a facility of the National Science Foundation operated under cooperative agreement by Associated Universities, Inc.} (termed the VLA-DIISC program), deep optical broad-band and H$\alpha$ imaging and spectroscopy with the MMT and VATT (termed the H$\alpha$-DIISC program), and combined them with archival ultraviolet and infrared imaging from the Galaxy Evolution Explorer (GALEX) and other space-based missions (termed as UV-DIISC program).

The goal of the DIISC survey is to investigate the multi-wavelength data concurrently to answer the following questions:
\begin{itemize}
\vspace{-0.2cm}
\item[1.] How are the gas content and kinematics of the CGM related to that of the \HI disk?
\vspace{-0.2cm}
\item[2.] How do the CGM and \HI disk's properties relate to the distribution of star-forming regions within galaxies?
\vspace{-0.2cm}
\item[3.] How are the metals distributed in the CGM, ISM, and stars?
\vspace{-0.2cm}
\item[4.] What are the dominant inflow and outflow pathways and processes that regulate star-formation in low-redshift galaxies?
\end{itemize}

In this paper, we describe the DIISC survey, including the sample selection, data acquisition, and reduction, and present some early results for the COS-DIISC program.

\section{Survey Design, Sample Selection, and Data Acquisition} \label{sec:sample}

To understand how the circumgalactic gas settles in the disk and subsequently cools into the neutral disk gas, we need observations that precisely measure the physical, chemical, and dynamical properties (including ionization state, temperature, metallicity, velocity dispersion, etc.) of the faint gas at the disk-CGM interface.
The faint, low-density nature of the disk-CGM interface region is best probed in absorption against bright background QSOs. QSO-absorption spectroscopy is sensitive to gas at column densities that are orders of magnitude lower than those detectable in any emission study with the current technology. It can also simultaneously probe multiple elements and their different ionization states and provide uniform signal-to-noise that is independent of the redshift of the targeted galaxy, allowing us to conduct a statistically unbiased study.

Intuitively, a good place to look for the gas in the disk-CGM interface would be beyond the 21cm \HI emitting denser \HI. Traditionally, the radius of the \HI disk, $\rm R_{HI}$, is defined at the surface mass density of gas of $\rm 1~M_{\odot}~pc^{-2}$ corresponding to an \HI column density of $\rm 1.3\times 10^{20}~cm^{-2}$. However, deeper imaging has conclusively shown that HI disks extend well beyond $\rm R_{HI}$ into a region commonly referred to as the extended disk \citep{Burstein81,bosma17, Wang13}.

Interestingly, the extended disk is the region where hydrogen transitions from optically thin to self-shielding. This critical phase change allows gas to survive ionization by the cosmic ultraviolet background.
The hydrogen column densities in the extended disk range from  $10^{17.2} <$ N(HI) $<10^{20.3}~\rm cm^{-2}$ \citep{kalberla09}.
Therefore, the region of interest to our program lies from the edge of the \HI disk ($ <R_{HI}$) through the extended disk ($ 1-2~R_{HI}$), out to the disk-CGM interface ($  2-\sim3~R_{HI}$). This region of the parameter space has yet to be fully explored by absorption surveys.

\subsection{Sample Selection}

\begin{figure*}[t]
\hspace{-.5cm}
\includegraphics[trim = 17mm 7mm 22mm 164mm, clip, angle=-0, width=3.6in]{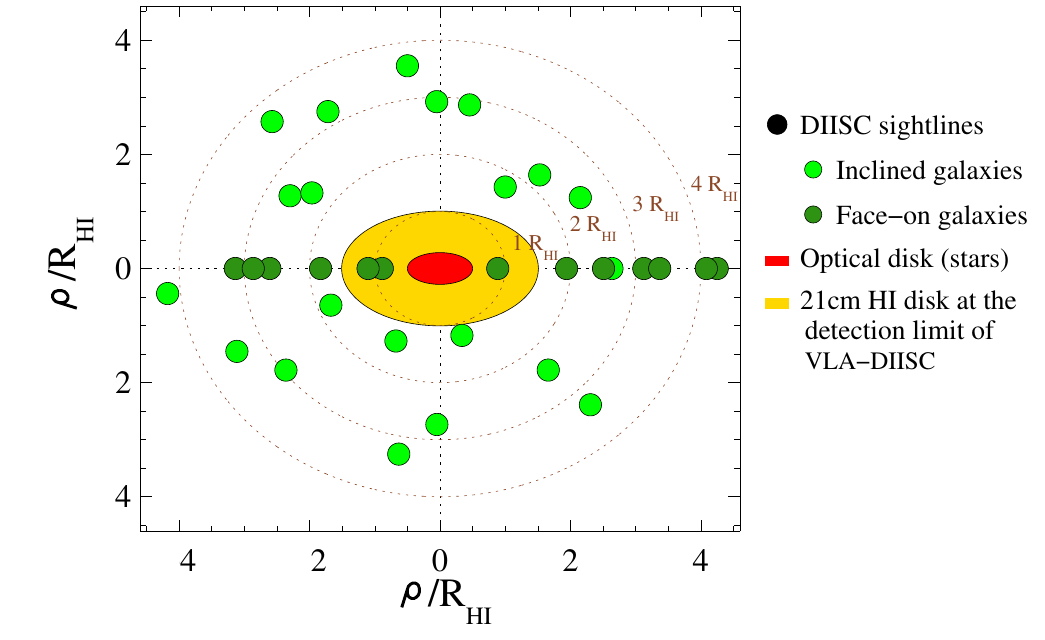}
\includegraphics[trim = 19mm 7mm 20mm 164mm, clip, angle=-0, width=3.6in]{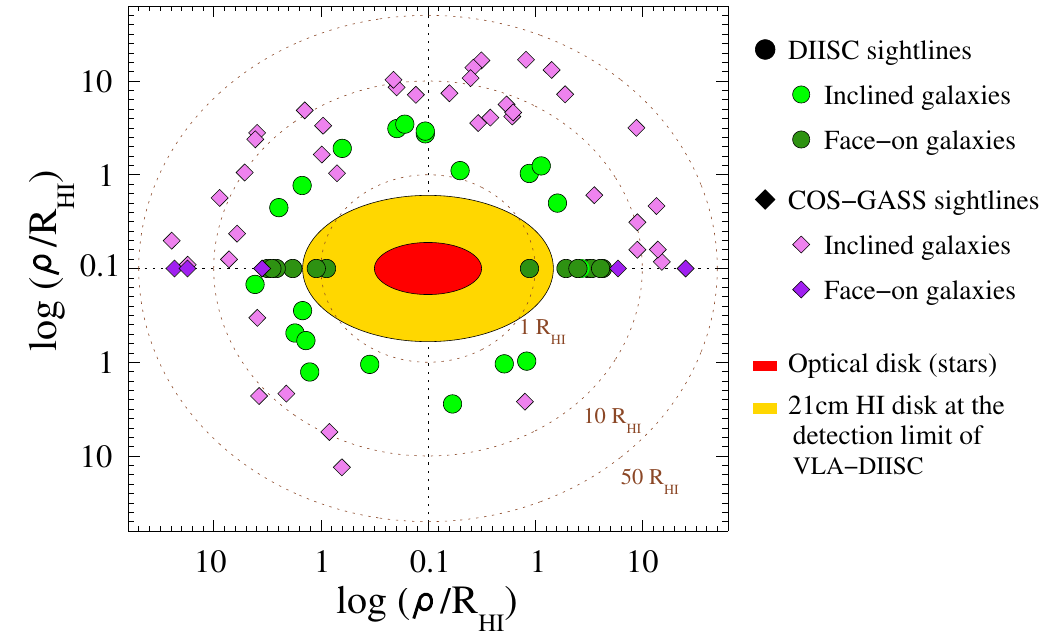}
\caption{ Overview of the QSO sightlines around DIISC galaxies. The figure shows a representative galaxy at the center, and the green points indicate the orientation of 34 QSO sightlines. The red and yellow ovals represent the approximate size of the stellar and gas disk, respectively. \HI\ 21cm imaging of the sample shows the \HI\ disk to extend down to $\sim~\rm 2 \times 10^{19} cm^{-2}$, about an order-of-magnitude fainter than used to define $\rm R_{HI}$. The left panel shows the DIISC sample, and the right panel shows the combined DIISC and COS-GASS samples on a logarithmic scale. The DIISC sightlines are light and dark green circles, whereas the COS-GASS sightlines are light and dark purple diamonds. The lighter and darker shades indicate face-on and inclined galaxies, respectively. The measurements for the GASS sample are adopted from the \citet{borthakur15}. Measurements for the DIISC sample are discussed in \S2.}
 \label{2D_schematic} 
\end{figure*}

 We identify a suitable sample by cross-correlating the  ALFALFA \citep{giovanelli05,haynes11} and HIPASS  \citep{Meyer04, Zwaan04} 21~cm \HI\ surveys with the catalog of unique GALEX GR5 QSOs \citep{bianchi11} to identify galaxy-QSO pairs.
The two \HI surveys, ALFALFA and HIPASS, have similar designs except for the depth and beam sizes. We use both as they probe different regions of the sky to increase the likelihood of finding a larger sample.

Following are our selection criteria that yielded an initial sample of 35 galaxy-QSO pairs.
\begin{itemize}
\vspace{-.2cm}
\item[(1)] The target galaxy is detected in \HI 21~cm emission;
\vspace{-.3cm}
\item[(2)] The impact parameter\footnote{projected distance of the QSO sightline from the center of the foreground galaxy at the rest-frame of the galaxy.} of the QSO sight-line is $\rm \rho \le 3.5~R_{HI}$ and $\rm \rho \le 200~kpc$ ;
\vspace{-.3cm}
\item[(3)] The apparent GALEX FUV magnitude of  the background QSO $ \rm m_{QSO}\le 19.0~mags$. This allows us to observe the targets with S/N of 10 per resolution element ($\sim \rm 15-20~km~s^{-1}$) in no more than 4 orbits with COS; and
\vspace{-.3cm}
\item[(4)] The difference in the redshift between the QSO and target galaxy, $z_{QSO} - z_{gal} > 0.02 (\equiv \rm 6000~km~s^{-1})$ to avoid confusion with absorbers associated with the QSO host. 
 \end{itemize}
 
Using the extremely tight relationship between \HI mass and size \citep[][and others]{Broeils97,swaters02}, we estimate the \HI disk sizes, $\rm R_{HI}$, from the \HI masses provided by the ALFALFA and HIPASS surveys. \citet{Stevens19} reports that the \HI size-mass relationship is also replicated in the Iluustris-TNG simulations and shows that this relationship is mathematically inevitable and robust analytically.
For galaxies where \HI imaging was available at the time of sample selection, we confirmed the accuracy of the predicted $R_{HI}$. However, high-resolution \HI\ 21cm observations with the VLA revealed that the \HI\ masses for seven galaxies are slightly smaller than those previously measured with the single-dish Arecibo telescope (more in \S~\ref{sec:vla_data}). The single-dish beam included nearby galaxies, mostly smaller (presumably companions). We adopted the new \HI masses from our VLA imaging, which resulted in smaller \HI radii for those galaxies. Consequently, five QSO-galaxy pairs now have impact parameters between $3.5 - 5~R_{HI}$, larger than the selection criteria. Nevertheless, we keep them in the sample as they nicely bridge between ours and the COS-GASS samples.

\begin{figure*}[t]
\includegraphics[trim = 10mm 110mm 50mm 80mm, clip, angle=-0, width=3.6in]{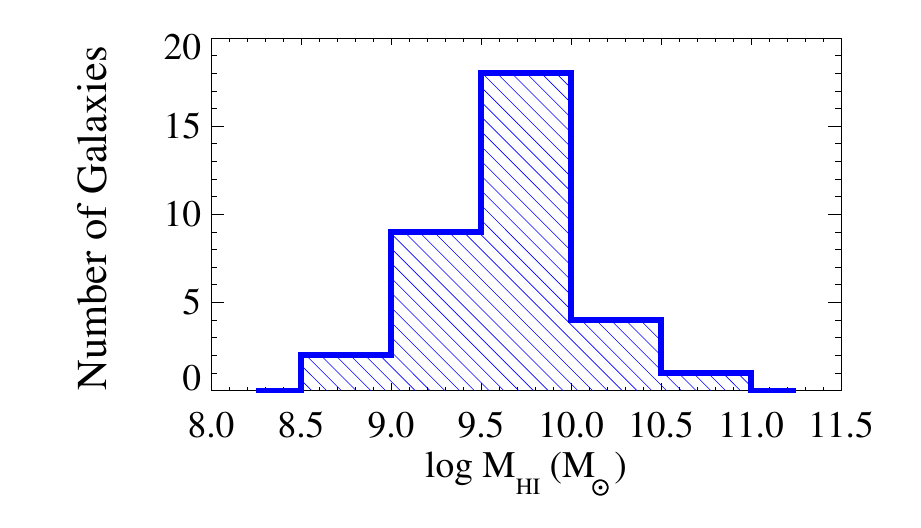}
\includegraphics[trim = 10mm 110mm 50mm 80mm, clip, angle=-0, width=3.6in]{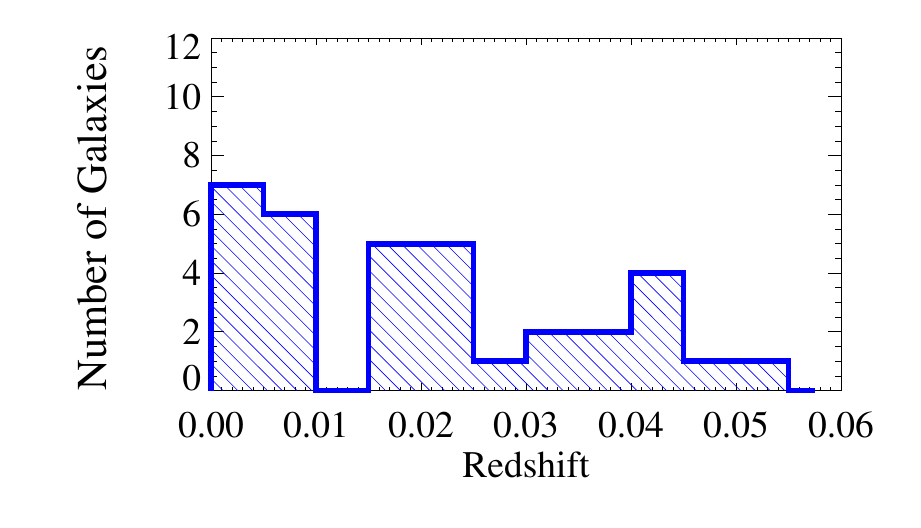}
\includegraphics[trim = 10mm 110mm 50mm 80mm, clip, angle=-0, width=3.6in]{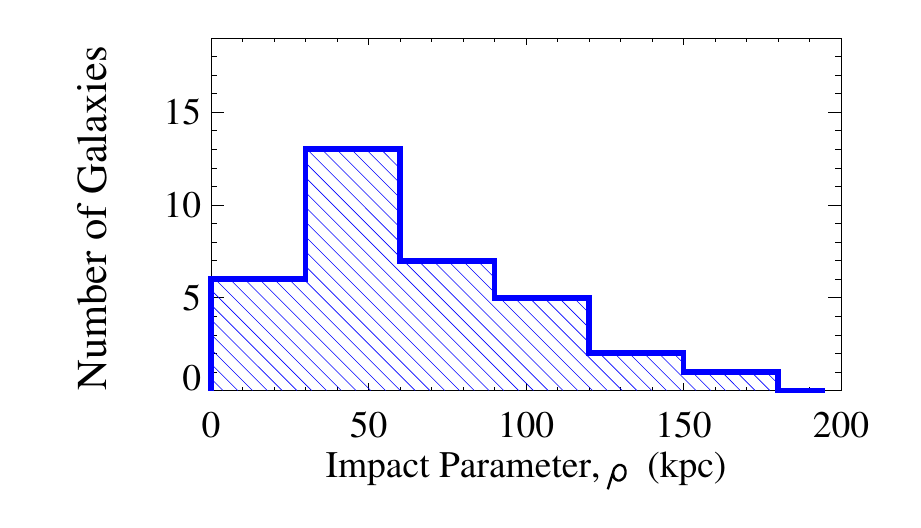}
\includegraphics[trim = 10mm 110mm 50mm 80mm, clip, angle=-0, width=3.6in]{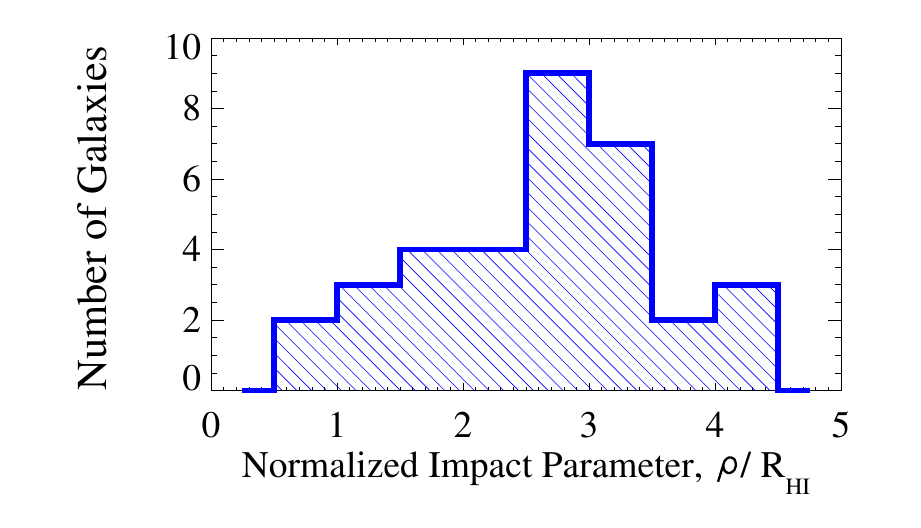}
\caption{Top left: Distribution of \HI\ 21cm mass of the DIISC sample. The \HI masses vary more than two orders of magnitude. The distribution peaks between $\rm log(M_{HI})$ of 9.5 and 10.0, which is comparable to \HI\ mass of the Milky Way \citep{kalberla09}. Top right: The distribution of redshifts of the DIISC galaxies. The redshift range is quite narrow owing to the accessibility of \HI\ 21cm line with ALFALFA. Since absorption line strengths are independent of redshift, the distribution of redshift of the targets does not impact the results presented in this paper. Bottom right: Distribution of impact parameter, $\rho$ for the DIISC sample. By design, we have limited the impact parameter to be $\le~\rm 200~kpc$. The Bottom left: Distribution of the impact parameter to \HI radius ($\rho/R_{HI}$). Although the sample selection criteria were set to $\rho/R_{HI} \le 3.5$, deeper imaging with the VLA resulted in newer, more accurate HI masses and radius for a few targets, thus leading to $\rho/R_{HI}$ to lie between 3.5--4.5 for five targets. }
 \label{HI_dist}
\end{figure*}

The distribution of the sightlines around our galaxies is shown in Figure~\ref{2D_schematic}. The left panel shows the sightlines from the DIISC sample around a representative galaxy, and the right panel compares the DIISC sample with the COS-GASS sample \citep{borthakur15}. The figures show a schematic of a representative galaxy, although each of our sightlines probes a different galaxy. 
The sightlines sample a wide range of orientations, including some face-on galaxies. When combined with COS-GASS, we will cover a large range of normalized impact parameters $\rm \rho/R_{HI}$. We also cover galaxies with \HI\ mass from $\rm log~M(HI)$ from 8.6 to 10.6 within a narrow redshift range of 0.002 to 0.051. The distribution of \HI\ masses and redshifts of galaxies are presented in Figure~\ref{HI_dist}.

We dropped one of the sightlines, J104709+130455, from the original sample as it probes one of the \HI structures associated with the Leo ring, not a galaxy matching our criteria. We would like to refer the reader to work by \citet{Sameer22} for analysis on that sightline.

\subsection{Data Acquisition}

We acquired multi-wavelength data covering the three phases probed in the survey, i.e., the CGM, ISM, and stars. We briefly describe data acquisition and reduction for our panchromatic data. Further details will be provided in future papers presenting those data.

\subsubsection{CGM traced by the COS}\label{sec:cos_data}

Our sample consists of 34 QSO-galaxy pairs probed by 32 QSO sight lines. The QSO sightlines were observed with the COS \citep{green12} aboard the HST from Nov 23, 2015, to Nov 4, 2016, under the program I.D. 14071. The observations were conducted with the G130M grating, giving a spectral resolution of R = 20,000--24,000, equivalent to 15--13 \kms. The integration time was chosen to reach a signal-to-noise of 10 for most parts of the spectral window. 
Our data covered rest-frame wavelengths, $\rm \lambda_{rest}= 1170-1430~\AA$, which gives us access to species such as  HI (\Lya),  \ion{O}{1} , \ion{C}{2}, \ion{C}{2}$^*$,  \ion{Si}{2}, 
\ion{Si}{3}, \ion{Si}{4},  \ion{N}{1}, and \ion{N}{5} for our sample. We also include archival data on some of our galaxies observed by previous HST programs 12248, 12025, and 13382, with the same grating at a similar signal-to-noise ratio. 

The spectra were created using the COS pipeline, calcos \citep{Soderblom22}. Absorption features in the spectra detected at a 3$\sigma$ level or higher were identified and matched to the species, transition, and redshift. The absorption line features include absorbers associated with the Milky Way, targeted galaxies from the DIISC sample, intervening systems, and the QSO host galaxies. We use a velocity window of  $\pm \rm 600$~\kms\ with respect to the systemic velocity (based on SDSS redshifts) to associate absorption features with our target galaxies. This velocity window is commonly used in other large CGM studies such as COS-Halos \citep{tumlinson13} and COS-GASS \citep{borthakur15,borthakur16}, thus making our analysis consistent with existing CGM studies.

Our continuum fits were performed, and equivalent widths were measured using the pipeline developed by the COS-Halos team \citep{tumlinson13, werk13}. 
For systems where the target absorbers are blended with Milky Way features, we modeled the Milky Way's Damped Lyman alpha and subtracted it to retrieve the absorption profile associated with the target. 
The modeling used the velocity centroid measured in \ion{Si}{2} by fitting multiple transitions covered in the data e.g. \ion{Si}{2} $\lambda\lambda$1190, 1193, 1260, 1304 $\rm \AA$. 
We used curve-fitting techniques similar to those of \citet{Tripp08} and \citet{tumlinson13}. We fitted Voigt profiles to the absorption features and derived the velocity centroids and b-values using the software by Fitzpatrick \& Spitzer (1997). The line-spread function of the COS G130M grating was folded into the fitting procedure. The number of components that were fitted was determined by visual inspection. For species with multiple transitions, the fit was derived by fitting all the transitions optimally. The H I column density and velocities were derived from \Lya alone, as we did not have access to any other transitions. The \Lya equivalent widths measured over all the components for the DIISC sample are presented in Table~\ref{tbl-W_LymanA}.  Metal lines will be presented in a later paper (Koplitz et al. in prep).

\subsubsection{ISM traced by the VLA}\label{sec:vla_data}

The DIISC galaxies were observed in the VLA D- and C-configurations under the large program 17A-090 and three regular programs, 18A-006, 19B-183, and 21A-408 for a total observing time of 282~hours. We acquired 6.5~hours of D-configuration data for each galaxy. In addition, we acquired 6.5~hours each of C-configuration data for the three most distant galaxies in order to resolve their disks. Some galaxies that suffered from radio-frequency interference (RFI) were re-observed in the later programs.

The observations were carried out with the 8-bit samplers to optimize the sensitivity of the spectral line observations. We used a total bandwidth of 16~MHz corresponding to a velocity coverage of 3500~\kms\ with 2048 channels translating to a channel width of 1.68~\kms. The wider velocity coverage is chosen to optimize the possibility of self-calibration using continuum sources in the field when available, in addition to looking for a wider range of companion/satellite galaxies. 
We observed galaxies at a spectral resolution of about 7.8~kHz (1.7 \kms) to minimize data corruption due to RFI. The calibrated data were then binned into various versions, including 20 \kms, comparable to the resolution of the UV data from the COS. The ability to image at multiple spectral resolutions allowed us to probe low-velocity dispersion structures such as anomalous velocity clouds \citep{gim21}.

The data reduction was performed in CASA 5.6.1 \citep{McMullin07} following the standard VLA data reduction, editing, and calibration steps. We produced moment 0, 1, and 2 images tracing the gas content, velocity, and velocity dispersion, respectively. The analysis of the VLA data and results will be presented in an upcoming paper by Gim et al. (in prep.). These data also produced broad-band radio continuum images of the galaxies.

\subsubsection{Young Stars traced by the GALEX Archival Imaging}\label{sec:galex_data}

We make use of archival far-ultraviolet (FUV) and near-ultraviolet (NUV) data from the Galaxy Evolution Explorer \citep[GALEX;][]{martin05,morrissey07}. These include data from surveys carried out by GALEX -- All-sky Imaging Survey and Medium Imaging Survey \citep[AIS and MIS, respectively;][]{morrissey07}, Nearby Galaxies Survey \citep[NGS;][]{bianchi03}, and the Guest Investigators Survey (GII). While AIS has typical exposures of $\sim100$~s, MIS, NGS, and GII have exposures $\sim1500$~s. The FUV and NUV images were produced by appropriately weighting and combining all available data for each galaxy and reached an average 1$\sigma$ sensitivity limit of 25.92~mag/\arcsec$^2$ in FUV and 26.02~mag/\arcsec$^2$ in NUV. The FWHM of FUV and NUV bands are 4\farcs2 and 5\farcs3, respectively, with a pixel scale of 1\farcs5/pixel.

\subsubsection{H~II regions traced via VATT H$\alpha$ Imaging}\label{sec:Halpha_data}

Deep narrow-band H$\alpha$ imaging was obtained from the 1.8 m Vatican Advanced Technology Telescope (VATT) operated by the Mt. Graham Observatory in Arizona. The observations were conducted over a total period of 19 nights between March 2019 to March 2021. The VATT4k CCD imager has a field of view of $\sim$12\farcm5$\times$12\farcm5 which comfortably fits all our galaxies in the sample and results in a pixel scale of 0\farcs375/pixel when the data is read-out with a $2\times2$ binning. 
We used seven different $\sim50$\AA\ wide, narrow-band H$\alpha$ filters from a wavelength range from 650~nm--700~nm to cover the redshifted H$\alpha$ emission line for galaxies effectively. Along with the H$\alpha$ imaging, we also secured deep broadband {\em r} imaging (in Sloan {\em r} filter) to allow subtraction of the stellar continuum.  Each galaxy had a total integration time ranging from 3600~s to 6000~s in the narrow-band filters to achieve H$\alpha$ fluxes as low as $\sim2\times10^{-16}$~erg~s$^{-1}$~cm$^{-2}$ even in the far outskirts. The total integration times in the {\em r}-band were $\sim1200$~s per galaxy. The typical seeing for these observations was between 0\farcs8--2\farcs0.

\subsubsection{Stellar distribution traced by optical broadband imaging}\label{sec:VATT_data}

The stellar distribution of the DIISC galaxies was traced using broadband {\em g} imaging (in Sloan {\em g} filter), also obtained from VATT during the observations described in \S\ref{sec:Halpha_data} along with the {\em r}-band imaging. The typical exposure times in {\em g}-band were about $600$~s and seeing between 0\farcs88--1\farcs80.

Furthermore, for part of the sample, we are securing deep optical broadband imaging from the $2\times8.4$~m  Large Binocular Telescope (LBT) using the Large Binocular Camera \citep[LBC;][]{giall08} to conduct a resolved stellar population study. The data is observed simultaneously in the Red and Blue channels of the LBC. We used the Bessel U, Sloan-{\em g} filter in LBC Blue and Sloan-{\em r} and Sloan-{\em i} filters in LBC Red. The observations seek exposure times to obtain surface brightness of $\sim27$~mag/\arcsec$^2$ at 5$\sigma$ in each filter and seeing $\lesssim1\farcs2$. The observations commenced in the Spring of 2021 and are currently ongoing.

\subsubsection{MMT/Binospec multiobject spectroscopy}\label{sec:Halpha_data}

We are conducting multi-object spectroscopy of \hii\ regions for a sub-sample of 14 nearby galaxies with the Binospec instrument on the MMT.  The spectroscopic data are obtained with the 270 lines/mm grating, providing a dispersion of 1.3\AA/pixel. 
The wavelength coverage of the data ranges from 3900--9240$\rm \AA$, making simultaneous measurements of \ha\ and other Balmer series lines, [OIII]\lamlam 4959, 5007, [NII]\lamlam6548, 6583, and [SII] \lamlam 6716, 6730, and multiple other nebular lines possible. We will also probe the [OII]\lamlam 3726, 3728 lines using a slightly different setting for a sub-set of galaxies.  
The program aims to investigate the metallicity distribution within these galaxies and identify signatures of recent low-metallicity gas accretion-driven star formation. Olvera et al. (in prep) will present results from the pilot program on NGC99.

\section{The Derived Measurements} \label{sec:derived_measuremnts}

We describe the estimation of all the derived quantities for galaxies. The values are listed in Table~\ref{tbl-samp}.

\subsection{Distance}
Since most galaxies in the sample are within 100~Mpc of the Milky Way, the closest being at 8.3~Mpc, Hubble flow distances derived from redshift measurement may have significant errors. Therefore, we adopt direct distance measurements whenever available or estimate distances from cosmic flow models. Table~\ref{tbl-samp} lists the distances and how they were derived.

The distances of the DIISC galaxies without direct measurements were estimated from the Extragalactic Distance Database \citep{edd2009} using their Numerical Action Methods (NAM) and Cosmicflows-3 (CF3) distance velocity calculators \citep{flowmodelcalc2020}. 
For galaxies with distances less than 38~Mpc, we use the NAM while the CF3 calculator was used to compute distances to galaxies with the observed velocities below $\sim15000$~km~s$^{-1}$ (or 200 Mpc, assuming the Hubble Constant of 75~km~s$^{-1}$~Mpc$^{-1}$).
The velocity field reconstructions for NAM and CF3 are based on \cite{shaya17}'s numerical action reconstruction and \citep{graz19}'s Wiener filter model for the peculiar velocity field, respectively.

\subsection{Galaxy inclination and orientation of the QSO sightline}\label{sec:prop}

We measured the position angles (PA) and axial ratios, $q=b/a$, where {\it a} and {\it b} are the semi-major and semi-minor axes, respectively, from our deep {\it r}-band maps using the publicly available Astropy package, \texttt{statmorph} \citep{statmorph}. The inclination angle, {\em i}, was then estimated following \cite{hubble26} as,
\begin{equation}\label{Eq-inclination}
Cos^2{i} = \frac{q^2-q_0^2}{1-q_0^2}
\end{equation}
where the intrinsic disk thickness, $q_0$, is assumed to be 0.2. The intrinsic thickness tends to vary between 0.15--0.25 and also depends on the galaxy morphology \citep{sandage70, aaronson80, fouque90} and stellar mass \citep{sanchez-janssen10, yu20}.  For most of our galaxies, $q_0=0.2$ would be applicable, although minor deviation from this value can not be ruled out. Taking this into account and measurement uncertainties, we estimate an uncertainty of no more than 10$^\circ$. The axial ratios, position angles, and inclinations are listed in Table~\ref{tbl-samp}.

The orientation of the QSO sightline was estimated based on the angle between the QSO and the major axis closest to the QSO. We consider galaxies with an axial ratio greater than 0.8 as face-on and those sightlines likely pass through the extended disk. An orientation value of 0$^{\circ}$ was assigned for those targets. The orientation of the QSO sightlines with respect to the galaxy is listed in Table~\ref{tbl-samp_qso}.

\subsection{Stellar mass and surface densities}\label{sec:stellar_mass}

Stellar mass maps for our galaxies were computed by combining VATT broadband imaging in the g- and r-band following the prescription by \citet{yang07}: 
\begin{equation}\label{Eq-Mstar}
\log\bigg[\frac{{\rm M}_{\star}}{h^{-2} {\rm M}_\odot}\bigg] = -0.406 +1.097(g-r) - 0.4\,({\rm M}_r - 5\log h -4.64)
\end{equation}
Here, M$_r$ is the absolute magnitude in \sdssr\ and (\sdssg\ - \sdssr) is the color derived from the \sdssg\ and \sdssr\ maps at each pixel. Pixels were masked if \sdssg\  - \sdssr~$>\pm3$ mag, which is a result of the subtraction of a source and source-free region in either map. 
The conversion of \mstar\ to \sm\ at each pixel was then carried out by using the scale values provided in Table \ref{tbl-samp} and pixel size of the VATT maps (0\farcs375).

The stellar masses derived from our optical photometry were compared to the masses derived from 2MASS Ks-band \citep{Munoz-Mateous07}. On average, our values were about 0.25~dex higher than the 2MASS measurements. Since there were no trends of a systematic offset, a calibration of the stellar masses couldn't be performed. In the rest of the paper, we note that there could be minor offsets from the derived relationships if an infrared-based stellar mass is used. 
As the primary requirement of the study is to acquire resolved maps and the 2MASS data is not suitable, we adopt the optical color-based map (Eq.~\ref{Eq-Mstar}) as our stellar mass.

\subsection{Halo mass and virial radii}\label{sec:Mhalo_Rvir}

The halo masses were derived from the stellar masses using prescription by \citet{Kravtsov18} and applying modifications based on the findings of \citet{mandelbaum16}. We follow the conversion used by \citet{liang14} and \citet{borthakur15} to estimate the virial radii using the relationship

\begin{equation}\label{Eq-Mhalo}
\rm R_{vir} = \frac{261}{(1+z)} \times M_{halo, 12}^{1/3}
\end{equation}
where $\rm M_{halo, 12}$ is the mass of the dark matter halo in units of $\rm 10^{12} M_{\odot}$ and $\rm R_{vir}$ has the units of kpc.

\subsection{Atomic hydrogen content}\label{sec:21cm_HI}

We estimate the surface density of atomic gas, \mhi, using the velocity-integrated flux density ($I_{tot}$) maps of the DIISC galaxies. At each pixel, \mhi\ is calculated using the relation from \cite{Verschuur_Kellermann88, mullan13}:
\begin{equation}\label{Eq-HI_mass}
    \Sigma_{\rm {H\,\small{I}}} = 1.0 \times 10^{4}\; \frac{I_{tot}}{\rm{A}_{\rm beam}}
\end{equation}
where $I_{tot}$ has units of Jy~beam$^{-1}$ km~s$^{-1}$ with the beam area, $A_{beam}$, expressed in squared arcsecond, and \mhi\ in M$_\odot$ pc$^{-2}$.

\subsection{Star formation rates and surface densities}\label{sec:sfr}

 The star formation rates are estimated using far-ultraviolet (FUV) data obtained by GALEX and 24$\mu$m infrared maps from Wide-Field Infrared Survey Explorer (WISE). These data were collected from archives and showed a large range in sensitivity as galaxies were covered at different depths. A few galaxies from the sample also have data in the Spitzer archives. However, to keep uniformity within the dataset for this analysis, we do not use the Spitzer maps; instead, we utilize WISE data.
 
 We derive the SFR surface density, \sfr, using the FUV and IR  prescriptions by \citet{salim07, calzetti07, leroy08} to estimate the \sfr\ at each pixel using the following relationships:
\begin{equation}\label{Eq-SFR1}
 \Sigma_{\rm SFR}\rm{(FUV+24\mu m)} = 0.081~{\rm I}_{\rm{FUV}} + 0.0032 ~{\rm I}_{\rm{24}} 
\end{equation}

where \sfr\ has units of M$_\odot$~yr$^{-1}$~kpc$^{-2}$ and I$_{\rm FUV}$ and I$_{\rm{24}}$  is in the units of MJy~sr$^{-1}$. Equation \ref{Eq-SFR1} assumes a 0.1--100 M$_\odot$ truncated Kroupa IMF \citep{kroupa01}. A detailed description of the procedure for creating the maps is provided in \cite{padave24}. The SFRs presented in this paper are integrated over the entire galaxy, defined as the region where FUV flux was detected. In certain cases where FUV or IR data is not available or extremely noisy (S/N $<$5), we opted to use IR and FUV measurements alone for estimating the SFR using prescriptions by \citet{cluver17}.
In the unique case, NGC~988, where both FUV and WISE data were noise, the SFR was estimated from 60$\mu$m flux adopted from \citet{Liu11} and the SFR prescription presented in \citet{Rosa-Gonzales02}.

\begin{figure*}[t]
\center 
\includegraphics[trim = 20mm 114mm 20mm 20mm, clip, angle=-0, width=3.5in]{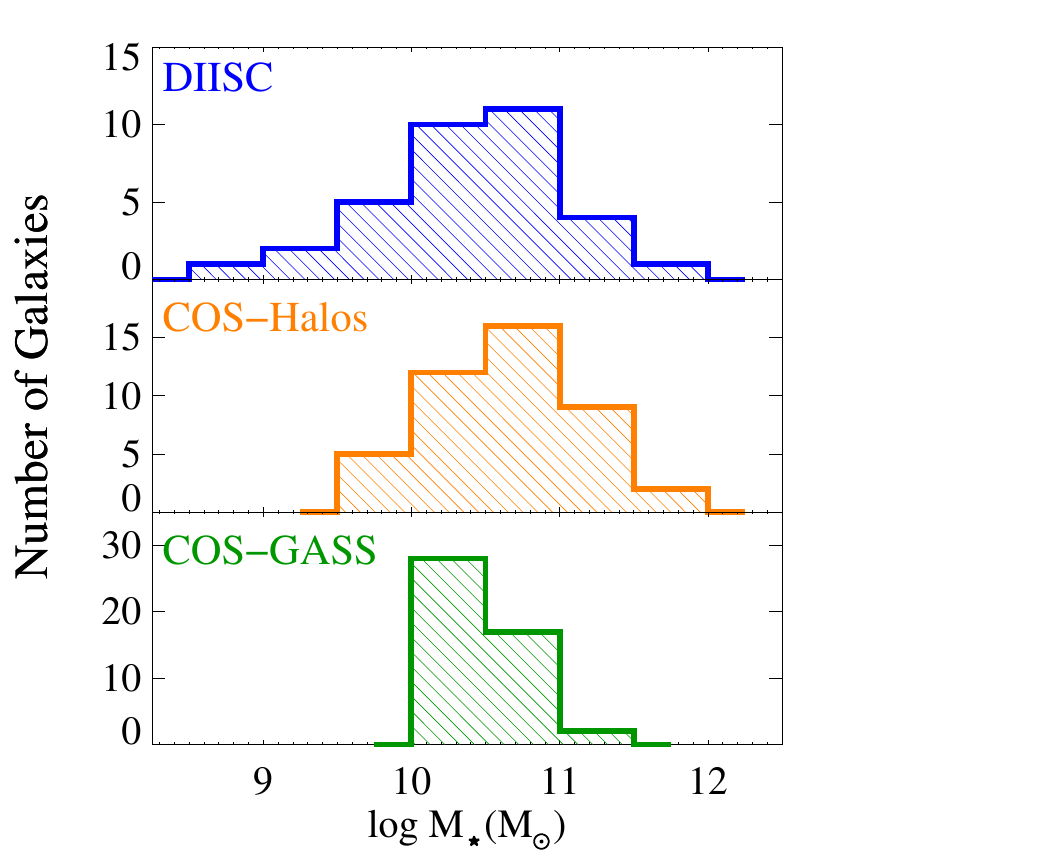}
\hspace{-1.5cm}
\includegraphics[trim = 20mm 114mm 20mm 20mm, clip, angle=-0, width=3.5in]{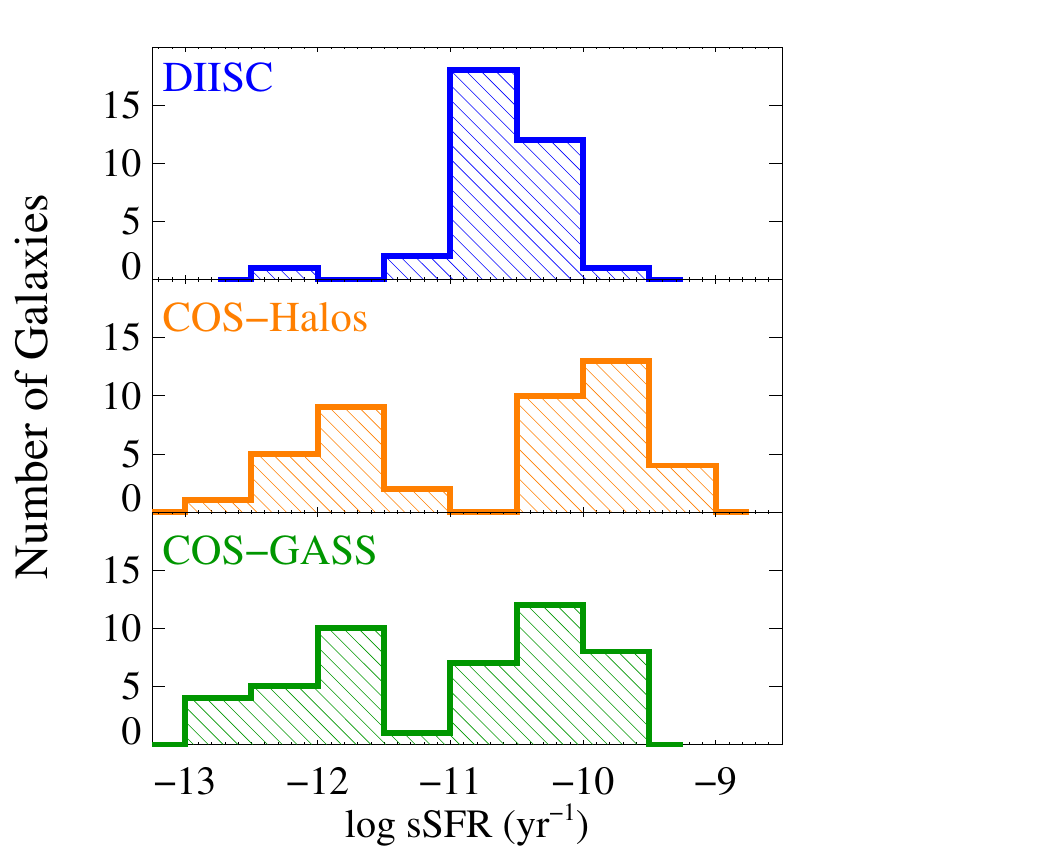}
\caption{Left: Distribution of stellar masses for the DIISC, COS-Halos, and COS-GASS samples. The DIISC sample spans a range of stellar mass from log~$\rm M_{\star} = 8.5-12$.  The distribution is very similar to that of the COS-Halos and COS-GASS samples, except for the low-mass tail. Right: Distribution of specific star formation rates for galaxies in DIISC, COS-Halos, and COS-GASS samples. Unlike stellar mass, the sSFR shows one prominent difference between the DIISC sample and both the COS-Halos and COS-GASS samples. Most of the DIISC galaxies cover the log sSFR$_{\{yr^{-1}\}}$ between $-$11 and $-$9.5 with a large fraction of green-valley galaxies \citep{martin07, salim07} unlike the COS-Halos and COS-GASS galaxies show a bimodal distribution in sSFR with only a few green-valley galaxies. This could be due to our selection bias towards large \HI disks. The peak of the sample matches with the sSFR of Milky Way of log~sSFR$_{\{\rm yr^{-1}\} } \sim -10.6 $ \citep{kalberla09, Elia22}.}
 \label{M_star} 
\end{figure*}

\section{Properties of the DIISC Galaxy Sample} \label{sec:properties_distribution}

The galaxy sample of our survey is similar to the two previous large HST-COS surveys, i.e., COS-Halos and COS-GASS. These surveys were selected based on different criteria. Nevertheless, the samples have good overlap. The COS-Halos sample was selected to have bright FUV bright QSOs (FUV mag $<$ 18.5) within $\sim$ a projected distance of 150~kpc of galaxies with photometric redshift  $z_{ph}>0.1$.  A consequence of the redshift criteria led to the selection of galaxies with stellar mass $\rm M_{\star} > 10^{10} ~M_{\odot}$. The COS-GASS galaxy sample was selected from the GALEX Arecibo SDSS Survey \citep[GASS; ]{catinella10} with background UV-bright QSOs with GALEX FUV mag $\le 19.0$~mag within an impact parameter of 250~kpc.

We present the distribution of the stellar masses of the DIISC sample in Figure~\ref{M_star} along with that of the COS-Halos and COS-GASS samples for comparison. The stellar mass of the sample peaks between $\rm 10^{10.5-11}~M_{\odot}$ and is similar to the COS-Halos sample with the exception that we have a few dwarf galaxies in the mass range $\rm 10^{8.5-9.5}~M_{\odot}$. The higher end of stellar mass distribution $\rm M> 10^{10}~M_{\odot}$  matches well with both the COS-Halos and COS-GASS samples.

\begin{figure*}[t]
\hspace{-1cm}
\includegraphics[trim = 00mm 114mm 0mm 80mm, clip, angle=-0, width=5in]{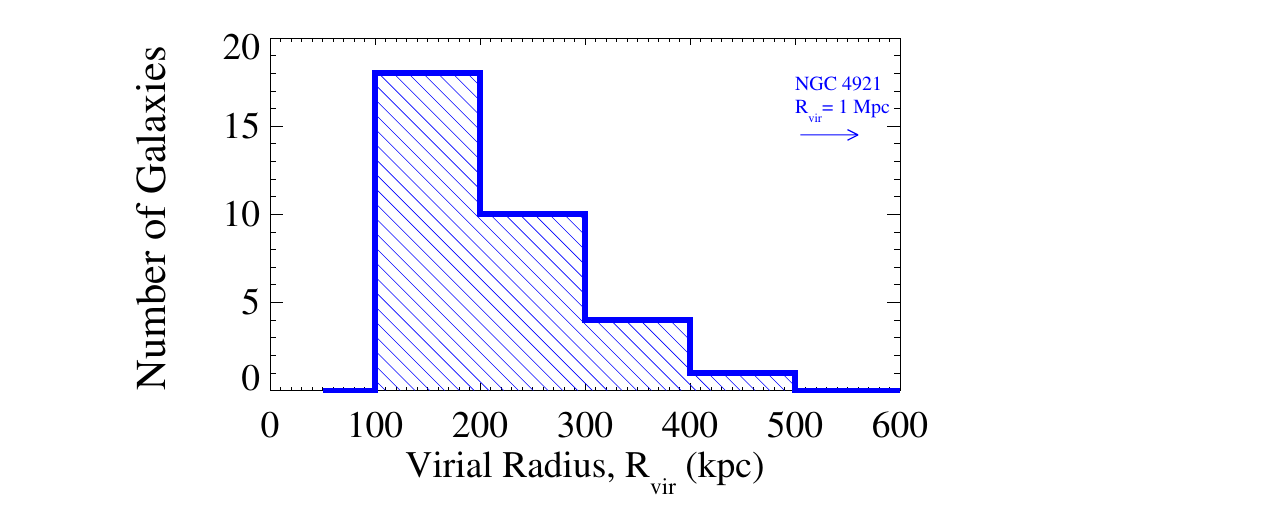}
\hspace{-3.5cm}
\includegraphics[trim = 0mm 114mm 0mm 80mm, clip, angle=-0, width=5in]{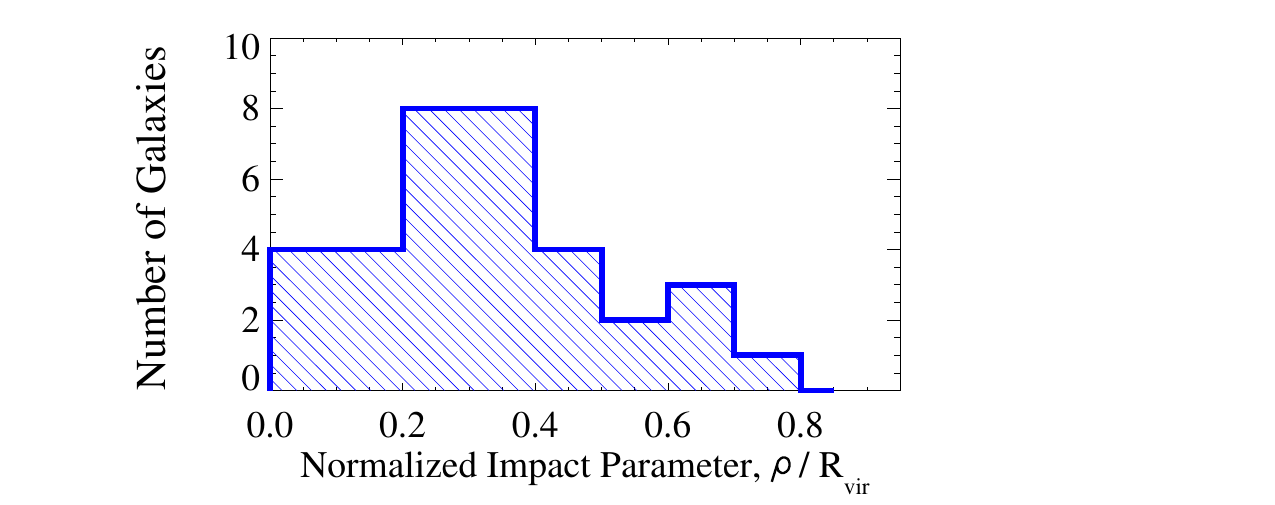}
\caption{Left: Distribution of virial radii for the DIISC sample. All but NGC~4921 and NGC~3728, have virial radii smaller than 350~kpc. NGC~4921, one of the prominent galaxies in the Coma cluster, is the most massive galaxy in our sample, with a virial radius of about 1~Mpc. Right: Distribution of the virial radius normalized impact parameter for the DIISC sample. The DIISC sample probes between 10\%-- 80\% of the halos of the galaxy. The large range of $\rm \rho/R_{vir}$  is a result of the unbiased nature of the sample in this parameter. }
 \label{Rho_and_Rvir} 
\end{figure*}

Similarly, the range of specific star formation rates of DIISC galaxies is broadly similar to the COS-Halos and COS-GASS samples. However, unlike the other two surveys, the DIISC sample does not go down to a very low sSFR. This is not surprising as this is an \HI\ selected sample. Owing to the strong correlation between atomic gas mass and SFR \citep{catinella10}, gas-rich galaxies are expected to be star-forming. The main difference, however, is that the DIISC sample does not show a bimodal distribution in sSFR, unlike the COS-Halos or the COS-GASS samples. This is most likely due to our requirement of having a background QSO that skewed the sample towards \HI-rich galaxies with larger disks. 
This may have resulted in a lack of red galaxies and an overabundance of green galaxies in the sample. 
Interestingly, most galaxies in the DIISC survey are like the Milky Way in terms of mass and SFR \citep{kalberla09, Elia22}. 
We label galaxies with log~sSFR$_{\{yr^{-1}\}}~ < - \rm 11$ as passive or red galaxies and the rest as blue galaxies. %

The distribution of the DIISC sample in terms of virial radii and normalized impact parameters are presented in Figure~\ref{Rho_and_Rvir}. Although the sample includes a large range of galaxy sizes, 90\% of the DIISC sample have virial radii of less than 350~kpc. The normalized impact parameter ($\rho/R_{vir}$) of sightlines peaks between 0.2--0.3 and probes the inner halo.

\begin{figure*}[!t]
\hspace{-1cm}
\includegraphics[trim = 5mm 114mm 0mm 80mm, clip, angle=-0, width=5in]{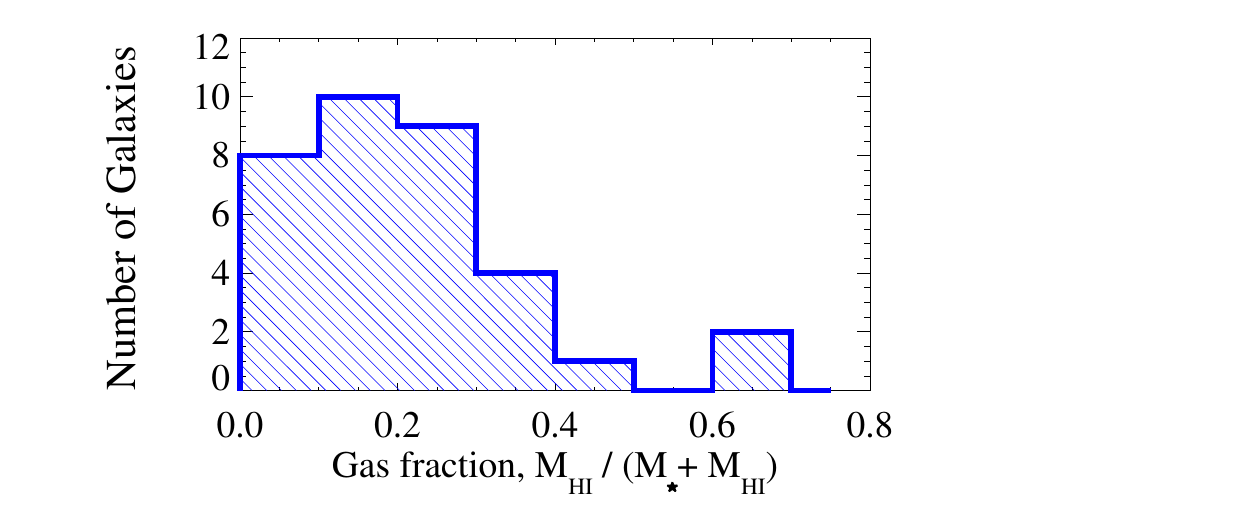}
\hspace{-3cm}
\includegraphics[trim = 5mm 114mm 0mm 80mm, clip, angle=-0, width=5in]{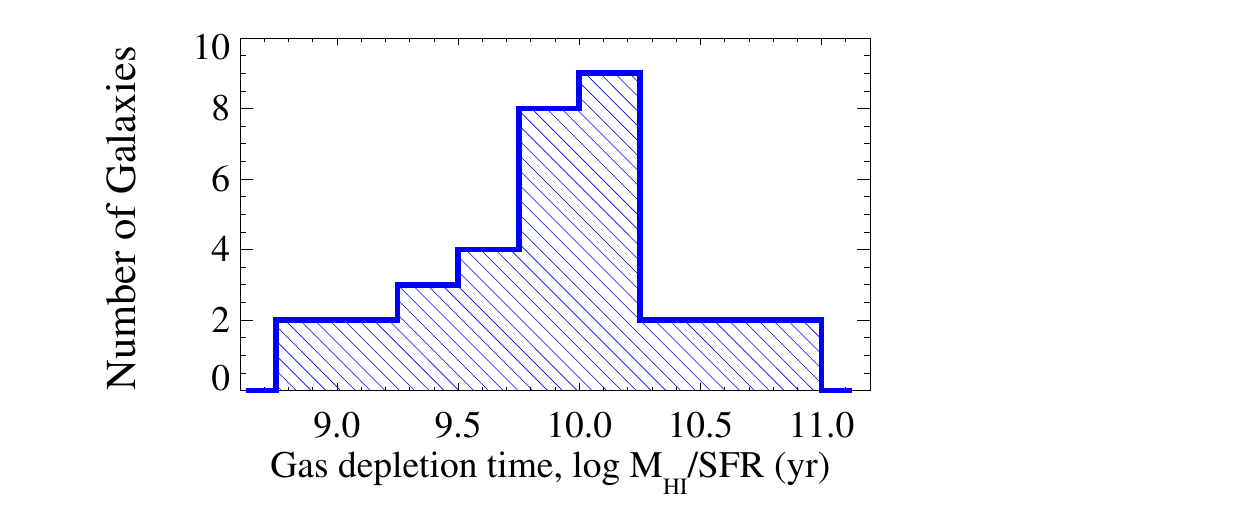} 
\hspace{-3.5cm}
\caption{Left: Distribution of gas fraction for the DIISC sample. The gas content ranges from 0.2\% of the total mass (stellar mass $+$ gas mass) to 63\%. Three-quarters of the sample have a gas fraction of at least 10\%. Right: Atomic gas depletion times ($\rm M_{HI}/SFR$) for the DIISC sample. The depletion times vary from a few Gyrs to hundreds of Gyrs.}
 \label{Gas_fraction} 
\end{figure*}

\begin{figure}[t]
\includegraphics[trim = 5mm 114mm 0mm 40mm, clip, angle=-0, width=4.2in]{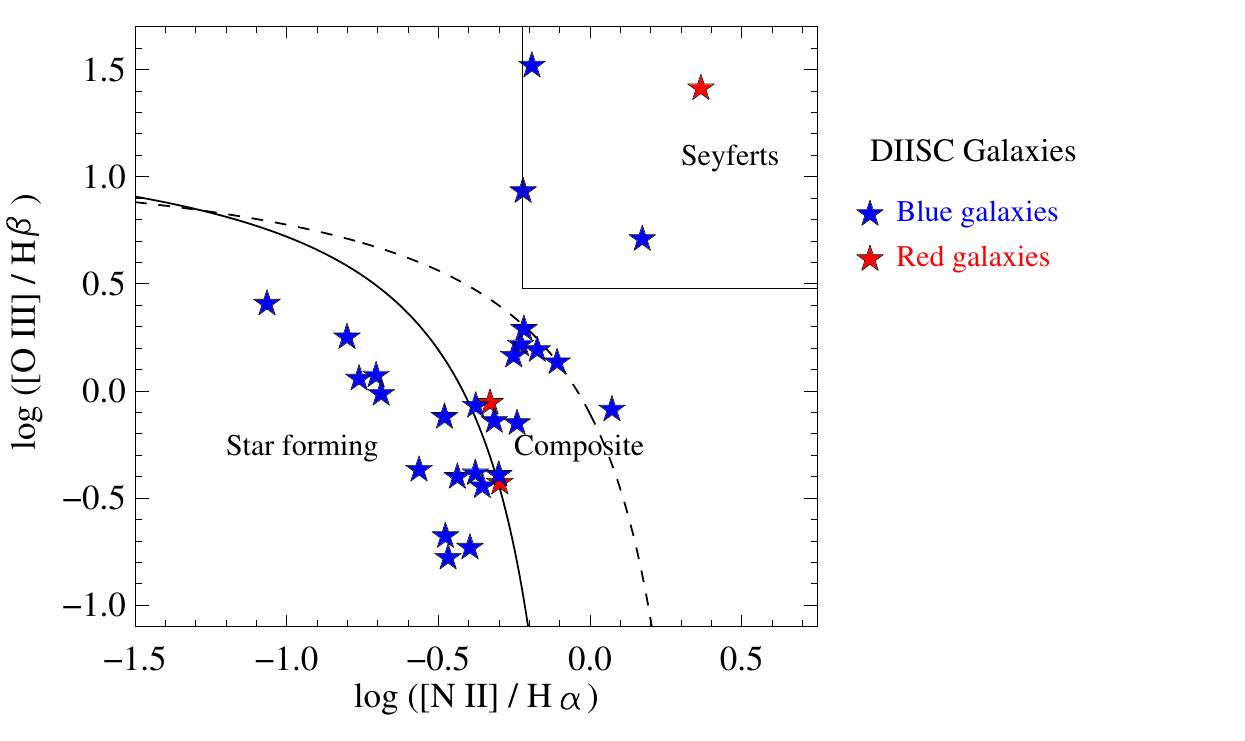}
\caption{DIISC galaxies on the Baldwin--Phillips--Terlevich (BPT) diagram. The emission line fluxes are obtained from SDSS DR12.  The solid and the dashed line show regions defined by \citet{kauffmann03} and \citet{kewley01}, respectively, distinguishing the regions where star-forming galaxies, star-formation AGN composites, and AGNs reside. Most DIISC galaxies lie in the star-forming region of the parameter space, a few in the composite region, and four in the Seyfert zone. Those are NGC~4921, IC1149, J0917+2720, and 2MASXJ1524+0421. Another four galaxies where H$\beta$ was not detected are omitted in this plot.}
 \label{BPT} 
\end{figure}

 The average gas fraction, $\rm M_{HI}/ (M_{\star}+M_{HI})$, of the DIISC galaxy sample is 0.21 with a minimum and maximum of 0.002 and 0.63, respectively. The distribution of the atomic gas fraction and the atomic gas depletion times ($\rm M_{HI}/SFR$) are shown in Figure~\ref{Gas_fraction}. The depletion times for the DIISC sample vary by two orders of magnitude from a few Gyrs to hundreds of Gyrs. None of the galaxies are active starbursts, and their sSFR ranges from  $\rm -12.05 \le log~ sSFR \le -9.96$. They also do not show signs of star formation-driven outflows in \ion{Na}{1} in the SDSS spectra that are traditionally seen in starburst galaxies. 
Except for four galaxies, most DIISC galaxies do not host a strong AGN. Figure~\ref{BPT} shows emission-line ratios from the nucleus of DIISC galaxies in the Baldwin--Phillips--Terlevich diagram \citep[][hereafter BPT]{BPT_81} that distinguishes type 2 AGNs from normal star-forming galaxies. The two lines mark the parameter space for the star-forming, AGN and composite galaxies defined by \citet{kauffmann03} and \citet{kewley01}, respectively.

%%%%%%%%%%%%%% RESULTS %%%%%%%%%%%%%%%%%%%

\section{Results} \label{sec:results}

The goal of this section is to investigate if there are correlations between the strength of the \Lya absorption line in the CGM and the bulk properties of the host galaxies. Here, we focus on the global properties of galaxies. Since the sample consists primarily of low-redshift galaxies, we have access to maps of these galaxies in terms of stellar and gas mass, SFR, and dust properties. However, a detailed analysis of the maps is left for the forthcoming paper.

\subsection{Nature of the circumgalactic medium at the disk-halo interface} \label{sec:CGM}

\begin{figure*}[]
\includegraphics[trim = 0mm 114mm 67mm 40mm, clip, angle=-0,height=2.6in]{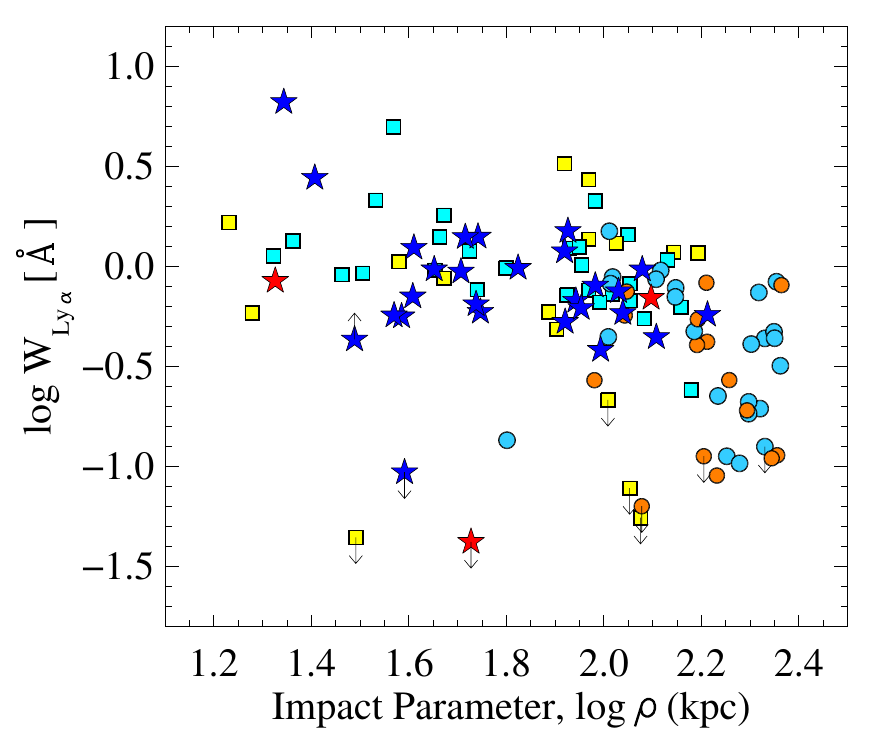}
\includegraphics[trim = 0mm 114mm 0mm 40mm, clip, angle=-0, height=2.6in]{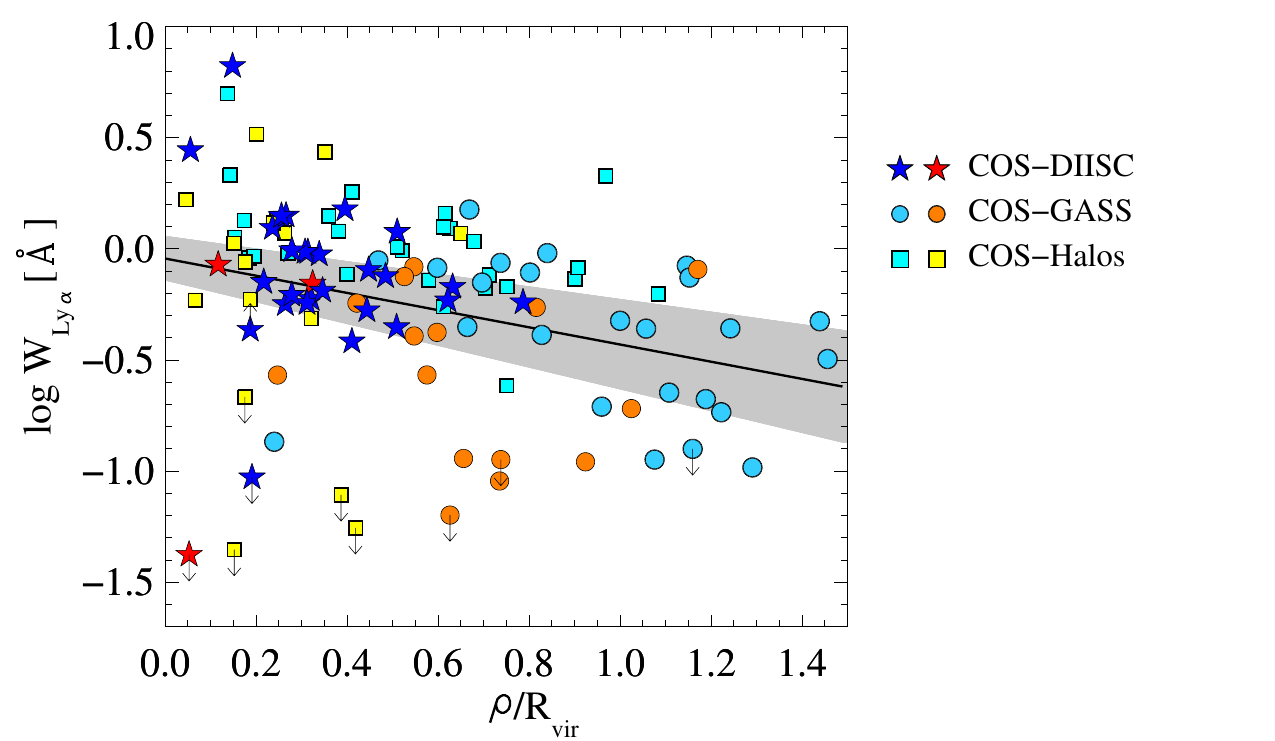}
\caption{Top Left: \Lya equivalent width as a function of impact parameter. The COS-DIISC sample is represented as stars, COS-Halos in squares, and COS-GASS as circles. The sSFR of galaxies is shown in color with blue, light blue, and cyan representing star-forming galaxies with $\rm log~sSFR_{\{yr^{-1}\}} \ge 10^{-11}$, and red, orange, and yellow representing passive galaxies with $\rm log~sSFR_{\{yr^{-1}\}} <-11$. The COS-DIISC shows a similar distribution as the COS-Halos and COS-GASS samples. The only non-detections are seen in the CGM of cluster galaxies-- NGC~4321 in the Virgo cluster and NGC~4921 in the Coma cluster. Top Right: \Lya equivalent width as a function of impact parameter normalized by the virial radius. The symbols and their color are the same as the left panel. The solid line shows the best fit to the COS-GASS and COS-Halos data combined published by \citet{borthakur16}. The gray-shaded region indicates the uncertainty in the fit. The COS-DIISC sample shows a statistically similar distribution to that of the COS-Halos and COS-GASS, although the sample selection criteria are entirely different. We used this best-fit line to define the \Lya profile as a function of the normalized impact parameter. The excess (or deficiency) of \Lya strength for any sightline is defined as the offset of the observed from this line (more in \S\ref{sec:discussion_Wexcess}).}
 \label{LyA_eqw_rho_Rvir}
\end{figure*}

The amount of neutral gas in the circumgalactic medium is known to be a strong function of the impact parameter of the QSO sightline \citep{chen98, morris06, steidel10, Tumlinson17}. Figure~\ref{LyA_eqw_rho_Rvir} shows the distribution of the \Lya equivalent widths for the DIISC sample (shown as filled stars; COS-DIISC hereafter) as a function of impact parameter ($\rho$) and impact parameter normalized by virial radius  ($\rho/ R_{vir}$). Data from the COS-Halos and COS-GASS surveys are overplotted as squares and circles, respectively. The color of the symbol represents the color of galaxies based on their sSFR, i.e., blue, light blue, and cyan represent galaxies with $\rm log~sSFR _{\{yr^{-1}\}}  \ge -11$, and red, orange, and yellow represent galaxies with $\rm log~sSFR _{\{yr^{-1}\}} < -11$.

The data indicate that the measurements for the DIISC sample are consistent with those of the COS-Halos and COS-GASS. The distribution with respect to impact parameter and virial radius normalized impact parameter do not show stark variations between the samples. This is interesting as the DIISC sample covers a larger range of stellar masses and was selected on entirely different criteria, i.e., proximity of the QSO sightline to the \HI\ disk, unlike the other two surveys that were selected on stellar mass and impact parameter of the QSO. The good match (within the scatter seen in both surveys) indicates that the CGM of the \HI\ galaxies is statistically indistinguishable from samples selected on stellar mass. The two diverging points from the COS-DIISC sample that fall far below the best-fit line are for sightlines passing through the CGM of NGC~4321 (M~100) and NGC~4921. Both these galaxies are part of galaxy clusters -- NGC~4321 is in the Virgo cluster, and NGC~4921 in the Coma cluster. The non-detection of \Lya in these galaxies is in agreement with the low \Lya covering fraction seen within the virial radius of cluster galaxies as was discovered by previous studies \citep{Yoon13, Yoon17, Burchett18}.

\begin{figure*}[t]
\includegraphics[trim =   0mm 114mm 20mm 43mm, clip, angle=-0, height=2.75in]{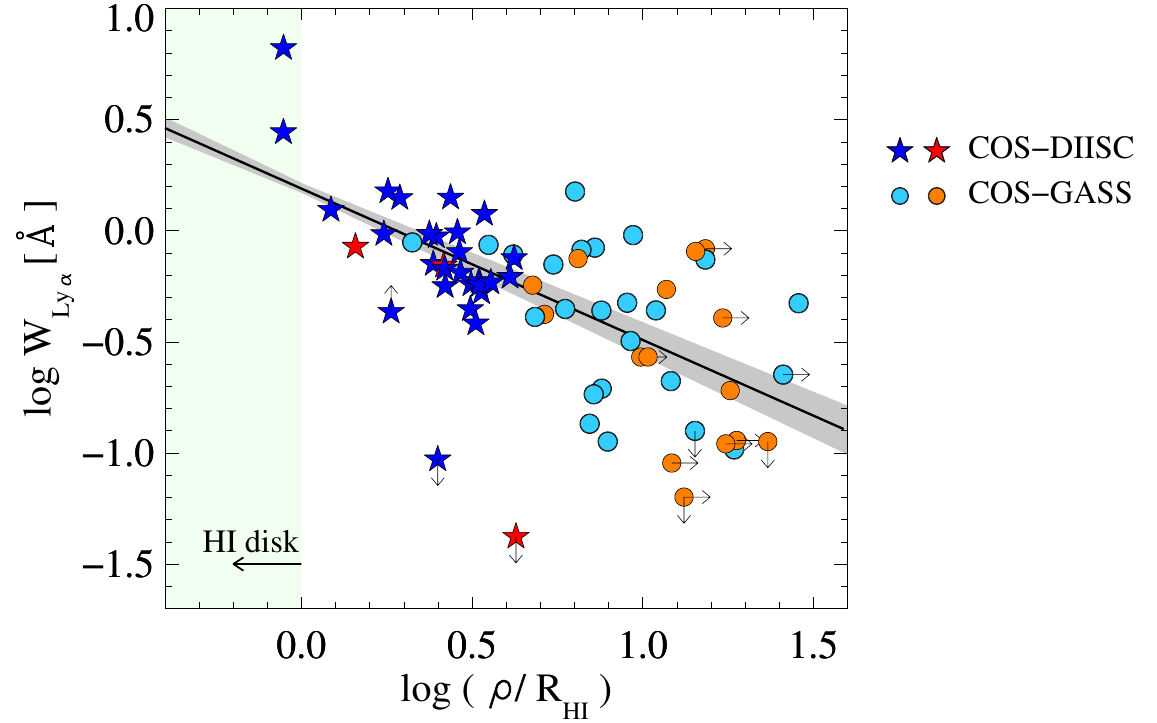}
\includegraphics[trim = 10mm 110mm 00mm 30mm, clip, angle=-0, height=2.75in]{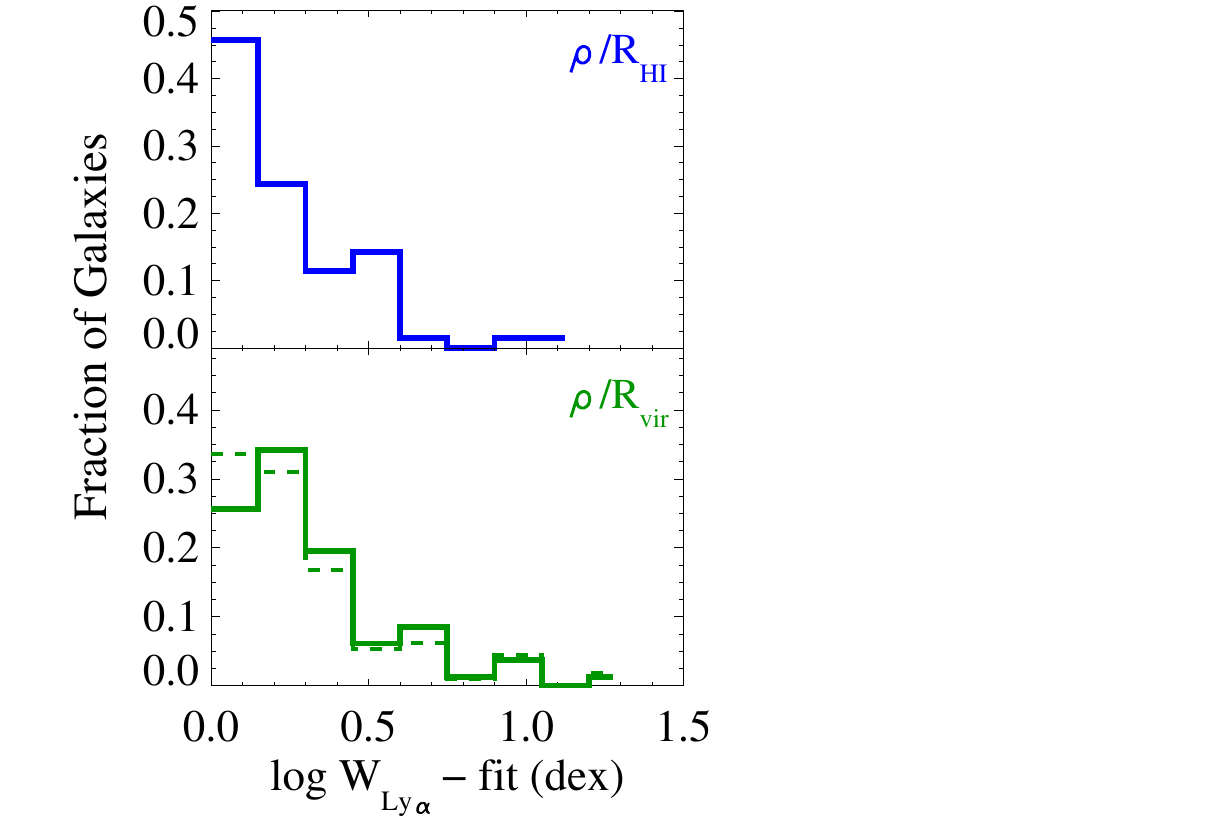}
\hspace{-3cm}
\caption{Left: Distribution of the \Lya\ content of COS-DIISC and COS-GASS sightlines as a function of distance from the edge of the \HI\ disk, i.e., impact parameter over \HI\ radius, $\rm \rho/R_{HI}$. The equivalent width shows a strong correlation at a confidence level of greater than 99.99\%. The best-fit line described in equation~\ref{Eq_rhoRHI} is shown as the solid black line and the 3$\sigma$ uncertainty in the fit (based on 1000 bootstrap iteration) is shown as the gray area. The symbols and their color represent the same as Figure~\ref{LyA_eqw_rho_Rvir}. Together, the two surveys cover more than an order of magnitude in terms of $\rm \rho/R_{HI}$, and the data shows a clear decreasing trend as a function of this parameter. The only two outliers are the non-detections from the two cluster galaxies, namely, NGC~4321 and NGC~4921. The light green shaded region indicates the \HI\ disk defined at a gas surface mass densities of 1~$\rm M_{\odot}pc^{-2}$ from \HI\ 21cm measurements. Right: A comparison of the distribution of the data shown in the left panel as well as Figure~\ref{LyA_eqw_rho_Rvir} with respect to the best-fit lines describing the relationship between \Lya equivalent width and \HI\ radius normalized impact parameter ($\rho/R_{HI}$) and virial radius normalized impact parameter ($\rho/R_{vir}$). The dispersion of the data in the case of $\rho/R_{HI}$ is 0.32 dex, with 45\% of the data lying within 0.15 dex of the fit. $\rho/R_{vir}$ plot shows a larger dispersion of 0.41(0.40)~dex with only 25(34)\% of the data lying within 0.15 dex of the fit for the COS-Halos+COS-GASS (COS-Halos+COS-GASS+COS-DIISC) samples shown as the solid (dashed) green line. Although the best-fit line for the data with respect to $\rho/R_{vir}$ was derived for the COS-Halos+COS-GASS sample, the addition of the COS-DIISC sample reduces the dispersion and elevates the fraction of data points within 0.15~dex of the fit. We also compared the dispersion of the combined sample from the three surveys to
In conclusion, we find that $\rho/R_{HI}$ is a better empirical predictor of \Lya equivalent width than $\rho/R_{vir}$.}
 \label{LyA_eqw_rhoRHI}
\end{figure*}

The availability of the 21cm~\HI\  measurements for the COS-DIISC and COS-GASS samples enables us to investigate the connection between the CGM and the \HI disk. Figure~\ref{LyA_eqw_rhoRHI} presents the \Lya equivalent width as a function of the distance from the \HI\ disk characterized by the ratio $\rho/R_{HI}$. 
Data from the COS-DIISC and COS-GASS are plotted as stars and circles, respectively, with colors (blue/red) indicating the log~sSFR  of the host galaxy above/below $-11~yr^{-1}$. The COS-DIISC sightlines probe the inner disk-CGM interface with $\rho/R_{HI}$ ranging from $\sim 1 - 4 $, whereas the COS-GASS sightlines fill the parameter space beyond $\rho/R_{HI}$ of 4 out to 30.  $\rho/R_{HI}$ shows much less scatter with equivalents width of \Lya than $\rho$ or $\rho/R_{vir}$ (see Figure~\ref{LyA_eqw_rho_Rvir}). This indicates that the CGM profile is more closely connected to the gas disk of the galaxy than to the stellar and dark matter content of the galaxy or a physical distance.

The data shows a strong inverse correlation at a confidence level of 99.99\% as estimated by the generalized Kendall's Tau test implemented in the survival analysis code, ASURV \citep{asurv}.
The \Lya equivalent width increases with decreasing $\rho/R_{HI}$ that can be best described	as:

\begin{equation} \label{Eq_rhoRHI}
\rm log~W_{Ly\alpha} = (-0.6810 \pm 0.0184) \times log~(\rho/R_{HI}) + ( 0.1894  \pm  0.0060)
\end{equation}
\noindent where, $ \rm W_{Ly\alpha}$ is the equivalent width of the \Lya absorption feature, $\rho$ is the impact parameter, and $\rm R_{HI}$ is the \HI\ radius of the galaxy defined using the relationship described by \citet{swaters02}. 
The fit was estimated using linear regression by Schmitt's method as implemented in ASURV. The uncertainties represent the dispersion in a bootstrap approximation of 1000 iterations. Due to limitations in the censored data methods, the data point with limits in both variables was considered as a limit on each variable separately.
The best-fit line is overplotted in Figure~\ref{LyA_eqw_rhoRHI} as the solid black line, and the 3$\sigma$ uncertainties in the fit, containing 99.7\% of the bootstrap iterations, are shown as the gray region. It is worth noting that the $\rm R_{HI}$ used in the analysis is estimated from $\rm M_{HI}$. Preliminary measurements $\rm R_{HI}$  from VLA \HI maps show a match with the relationship. We will present $\rm R_{HI}$ measurements from our VLA imaging in a future paper.

The dispersion in the relationship described in Eq.~\ref{Eq_rhoRHI} is mostly seen at larger distances from the disk. Both red and blue galaxies show similar dispersions. We note that this relationship is probably only valid for the region explored in the sample and might break down as we enter the denser regime within the \HI disk at mass surface densities higher than 1$\rm M_{\odot}~pc^{-2}$. A hint of this can be seen with the two COS-DIISC sightlines that probe within the disk and show much larger equivalent widths. Although those points were used to generate the fit, it is likely that the gas physics changes at those column densities. The gas becomes self-shielding and may support much larger neutral \HI\ column densities. The two \Lya non-detections in COS-DIISC correspond to NGC~4921 and NGC~4321. They have the highest offset from the best-fit line. This is likely the result of the hot cluster media impacting the CGM of these galaxies. Corroborating evidence of impact on the ISM of NGC~4921 was seen in our VLA \HI 21cm imaging that shows strong signs of ram-pressure stripping, \HI asymmetry, and deficiency consistent with previous studies \citep{Kenney15, Cramer21}.

The distribution in the \Lya equivalent width offsets from the $\rho/R_{HI}$ fit derived in Eq~\ref{Eq_rhoRHI} is shown in the right panel of Figure~\ref{LyA_eqw_rhoRHI}. For comparison, we also show the distribution of the offsets from the best fit derived from $\rho/R_{vir}$ by \citet{borthakur15}. The dispersion of the data with respect to the best fit is lower for $\rho/R_{HI}$ with 45\% of the points lying within 0.15~dex of the fit. Meanwhile, the fit for $\rho/R_{vir}$ has a higher dispersion %of 0.41~dex 
with only 25\% of the data within 0.15~dex of the fit.
We compared the spread in the offset for $\rho/R_{vir}$ with or without the COS-DIISC points. 
The dispersion in the data with respect to the fit shows a minute drop of 0.01 dex (statistically insignificant) %to 0.40~dex 
when the COS-DIISC sightlines are added to the sample.
The fraction of points within 0.15~dex increases to 34\%, suggesting that the DIISC sample is consistent with the expectations derived from the previous surveys.
We also compared the data to the best-fit plane described by \citet{bordoloi18} that relates equivalent width to $\rho$ and $\rm M_{\star}$. We found the offsets to have a dispersion of 0.8~dex, greater than those found for $\rho/R_{HI}$ and $\rho/R_{vir}$.

In conclusion, the tightness of the fit indicates that $\rho/R_{HI}$ is more robust in predicting \Lya strength in the CGM compared to the traditionally used parameters, $\rho/R_{vir}$. This suggests that the gas distribution in the CGM is more closely related to the gas disk than the stellar or dark matter component in local galaxies.

%%%%%%%%%%%%%%%%%%%%%%%%%%%%%%%%%%%%%%%%%
\section{Discussion}\label{sec:rdiscussion}

\subsection{Connecting the CGM to the \HI disks of galaxies }

The gas distribution in the CGM is believed to be closely related to the gas flows that sustain the gas disks in galaxies. 
In this context, the observed tight relationship between the equivalent width of \Lya and the proximity of the sightline to the disk is not surprising.

A recent stacking study of \HI maps by \citet{Ianjamasimanana18} showed that disks gradually fade to lower column densities and do not show a break at about $\rm 5\times 10^{19}~cm^{-2}$ unlike previously thought \citep{Corbelli93}. These authors detected gas disks down to the sensitivity of their data at a few times $\rm 10^{18}~cm^{-2}$.
Interpreting our observations in the context of their result suggests that the extended gas disks go down to much lower column densities and can still be traced in absorption,

Absorption studies probing \ion{Mg}{2} as well as other absorbers have shown evidence of co-rotation, i.e., the same velocity direction as the side of the disk near the sightline \citep{Kacprzak10, Diamond-Stanic16, Ho17, martin19, French20, Dupuis21, Nateghi23a, Nateghi23b}. In some cases, excellent velocity matches between the CGM and the stellar disks were seen, while in some cases, anti-direction has also been seen \citep{Kacprzak10, Ho17}.
However, these measurements do not utilize the rotating gas disk but the stellar disk, and as a result, mild divergence in kinematics from the stellar disk is hard to interpret. 

Multiple simulations also predict that the outer gas disks should transition into the circumgalactic gas at larger radii. However, the specific properties of this region differ between simulations.
Simulations by \citet[][and others]{Stewart11, Nuza14, Ho19} found large rotating disks in low-redshift galaxies similar to the ones in the DIISC sample. 
 \citet{Rottgers20} found the gas disk to become fuzzier into a diffuse CGM that drops in \Lya equivalent width with galactocentric distance. Recently, \citet{hafen22} in the FIRE simulations find a unique stage in the gas accretion process that builds up a large co-rotating disk-CGM interface.
In their simulations, the angular momentum of accreting warmer gas was aligned to the disk, which resulted in the gas cooling into a flattened disk-like configuration.

In general, simulations predict that the inner CGM at the disk-CGM interface has some continuity with the extended \HI disk \citep{ramesh23}. However, quantitative comparisons do show variations between the observed and the predicted.  
For example, a comparison of Illustris simulations and the COS-GASS survey by \citet{Kauffmann16, kauffmann19} found that those simulations under-predicted the covering fraction seen in the observations. Although the precise reason for this offset is not certain, further comparisons incorporating the new data from COS-DIISC and newer simulations hold the promise to provide valuable insights into the process of disk formation and growth.

\begin{figure*}[t]
\includegraphics[trim = 0mm 114mm 67mm 40mm, clip, angle=-0, height=2.5in]{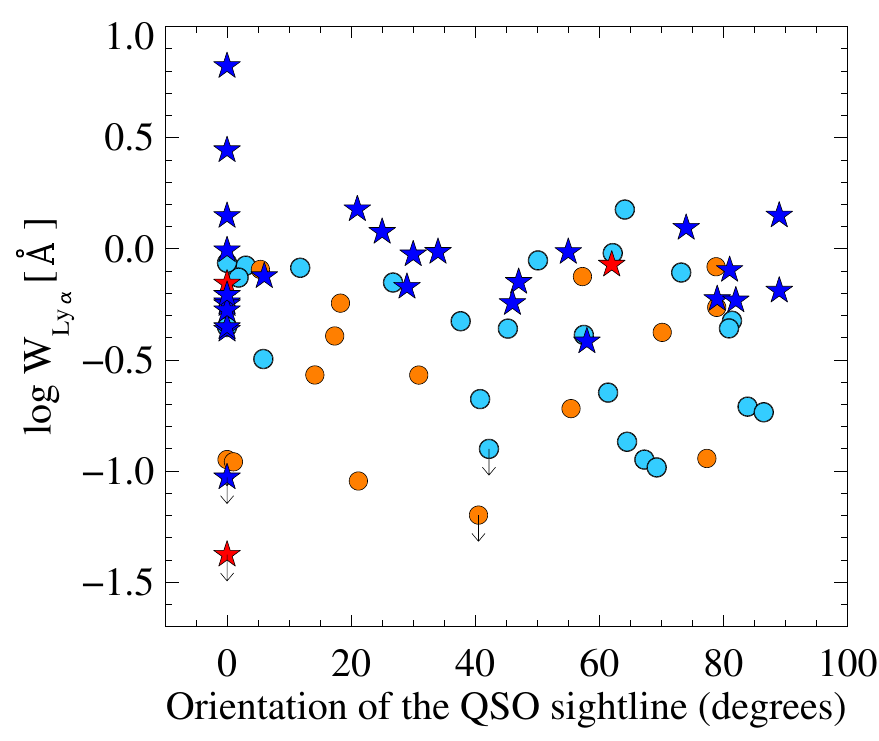}
\hspace{0.5cm}
\includegraphics[trim = 0mm 114mm 0mm 40mm, clip, angle=-0, height=2.5in]{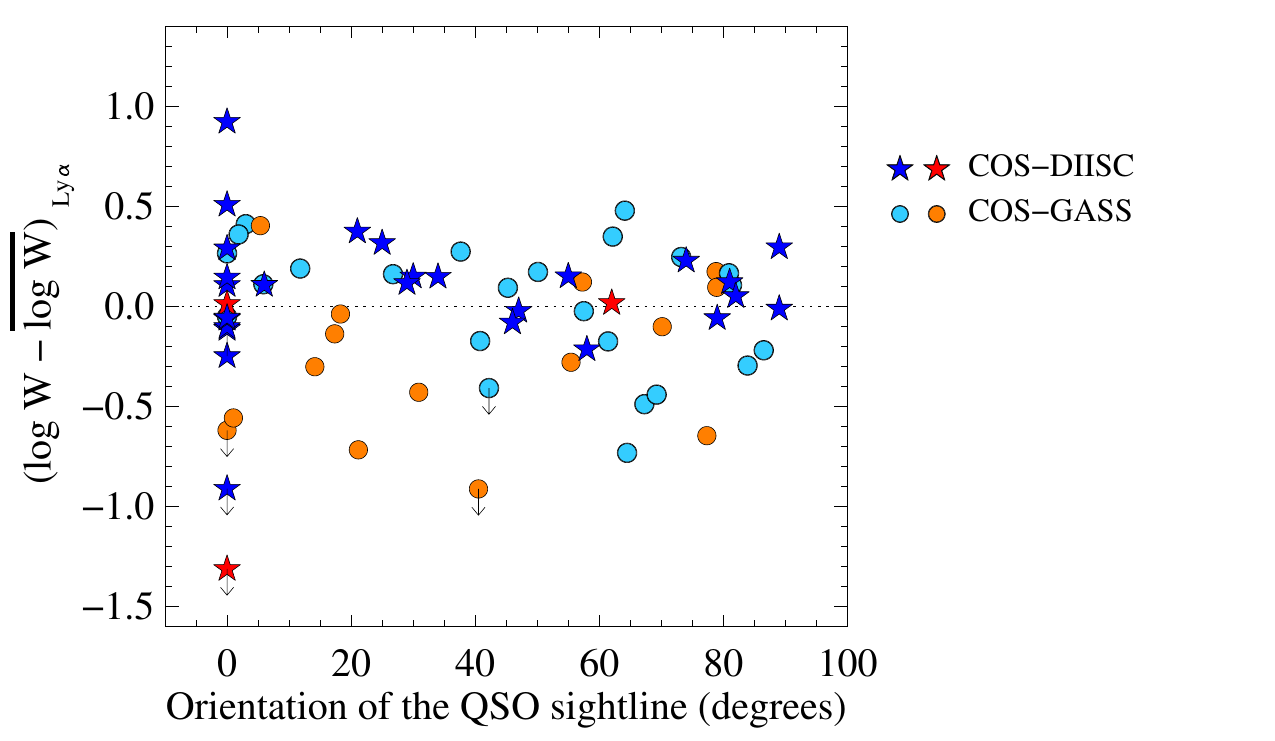}
\caption{\Lya equivalent width (left panel) and excess in the \Lya equivalent width, $\rm log~W_{Ly\alpha}^{excess} = log~W_{Ly\alpha} - \overline{log W}$, (right panel) as a function of the orientation of the QSO sightline with respect to the disk. The color of the symbols represents their sSFR with blue and light blue for galaxies with log~sSFR$_{\{yr^{-1}\}} >-11$, and red and orange for galaxies with log~sSFR $<-11$. The shapes of the symbols are indicated in the legends. We see no correlation between the orientation of the QSO sightline and the neutral \HI content in their CGM.}
 \label{Wexcess_orientation}
\end{figure*}

While our result indicates that a substantial fraction of the \Lya absorption is likely caused by circumgalactic gas that is somehow physically connected to the \HI disk, the morphology of the circumgalactic gas beyond the 21cm \HI disk in the disk-CGM interface need not be a well-defined disk. 
Instead, the extended disk will likely flare up and become more volume-filling \citep{Vollmer16, bosma17}.
Deep imaging studies in \HI 21cm transition have shown that the disks flare up at large galactocentric radii \citep{Vollmer16, Das19}, although it is worth noting that these regions are orders of magnitudes higher in column densities that our sample is probing. Nevertheless, the idea of a flaring disk is consistent with the fact that we do not see a dependence between \Lya equivalent width and the orientation of the QSO with respect to the disk (Figure~\ref{Wexcess_orientation}). 
The absence of correlation may indicate that the extended gas disk at the disk-CGM interface is a flared disk. This may wipe out any orientation effects for slightly inclined or face-on galaxies, i.e., sightline passing closer to the disk may intersect the disk due to its inclination.
For instance, a low column density disk of about 100~kpc (similar to the predictions by \citet{Stewart11}) at an inclination of 45$^{\circ}$ would have a projection 50~kpc in the vertical direction.  A perfect sample of edge-on disks would be required to check for this, and it is currently beyond the scope of the DIISC sample.

Since the $\rm R_{HI}$ and $\rm M_{HI}$ are directly (one-to-one) related, the correlation between \Lya equivalent width and $\rm \rho/M_{HI}$ ($\rho/M_{HI}$) is also strong.  However, we argue that the $\rm R_{HI}$ is likely the principal variable that determines the neutral gas content of the CGM as it provides a physical connection between the gas in the disk and the CGM. The \HI radius very well characterizes the maximum size of the entire gas disk, including the molecular component. Meanwhile, the total gas mass in the disk is not well characterized by the \HI mass. The atomic gas transitions into molecular form in the inner disk, and the amount of molecular gas to atomic gas varies among galaxies \citep{young_scoville91}. The gas disk is almost always as big as the \HI disk, but the total gas mass can be significantly larger than the \HI mass. Using published molecular gas mass for nine of the DIISC galaxies, we investigated the correlation between \Lya equivalent width and the total gas mass, i..e, $M_{HI+H_2}$, and impact parameter over total gas mass ($\rho/M_{HI+H_2}$) and found no significant correlation.

\subsection{Investigating the role of bulk galaxy properties impacting the CGM radial profiles  } \label{sec:discussion_Wexcess}

The relationship between \Lya equivalent width and normalized impact parameter ($\rho/R_ {vir}$) shows a large scatter, which suggests that there could be other variables (galaxy properties) that influence the strength of the \Lya equivalent width.
We investigate how the scatter is related to some of the common bulk properties like \HI\ mass, gas fraction, SFR, and sSFR. We define the excess (or deficiency) in the \Lya equivalent width, $\rm W_{Ly\alpha}^{excess}$, by calculating the offset of the observed value from the best-fit line relating $\rm W_{Ly\alpha}$ and $\rho/\rm R_{vir}$. The best-fit line, also shown in the right panel of Figure~\ref{LyA_eqw_rho_Rvir}, was derived by \citet{borthakur16} from the COS-Halos and COS-GASS data. Mathematically,

\vspace{-0.5cm}
\begin{equation} \label{Eq_Wexcess}
%\rm log~W_{Ly\alpha}^{excess} = log~W_{Ly\alpha} - \overline{log W} =  log~W_{Ly\alpha}  - (-0.387~(\rho/R_{vir}) - 0.044 )
\rm log~W_{Ly\alpha}^{excess} = log~W_{Ly\alpha}  - (-0.387~(\rho/R_{vir}) - 0.044 )
\end{equation}

\begin{figure*}[t]
\includegraphics[trim = 0mm 114mm 67mm 40mm, clip, angle=-0, height=2.5in]{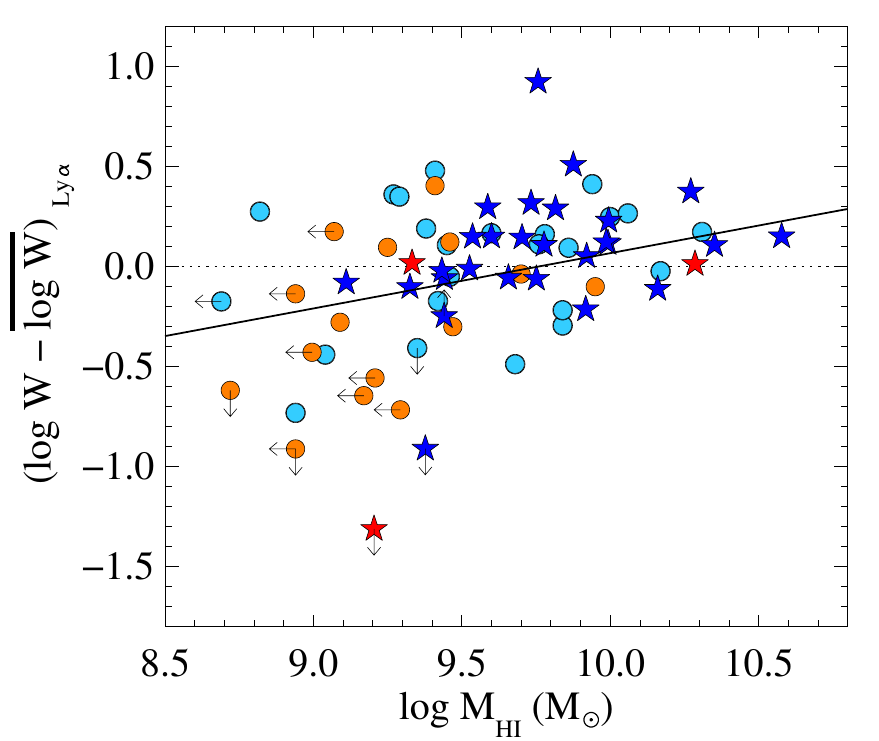}
\hspace{0.5cm}
\includegraphics[trim = 0mm 114mm 00mm 40mm, clip, angle=-0, height=2.5in]{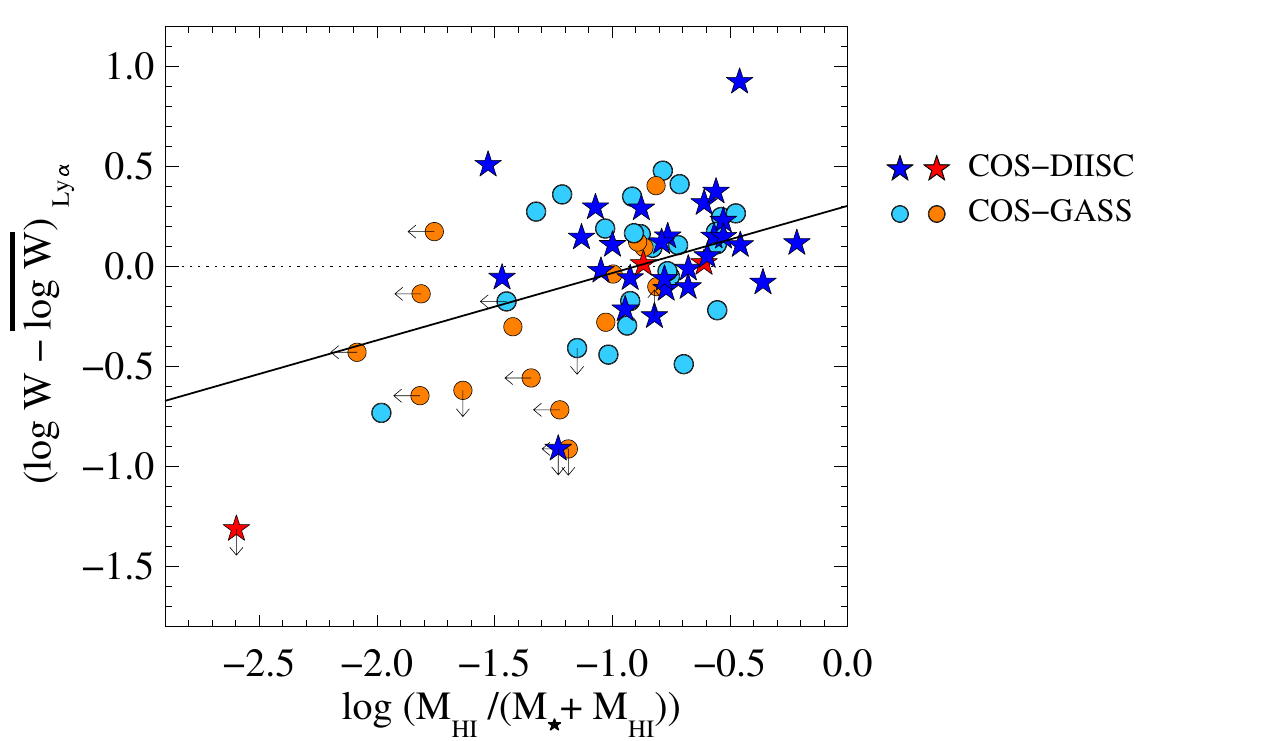}

\vspace{0.5cm}
\includegraphics[trim = 0mm 114mm 67mm 40mm, clip, angle=-0, height=2.5in]{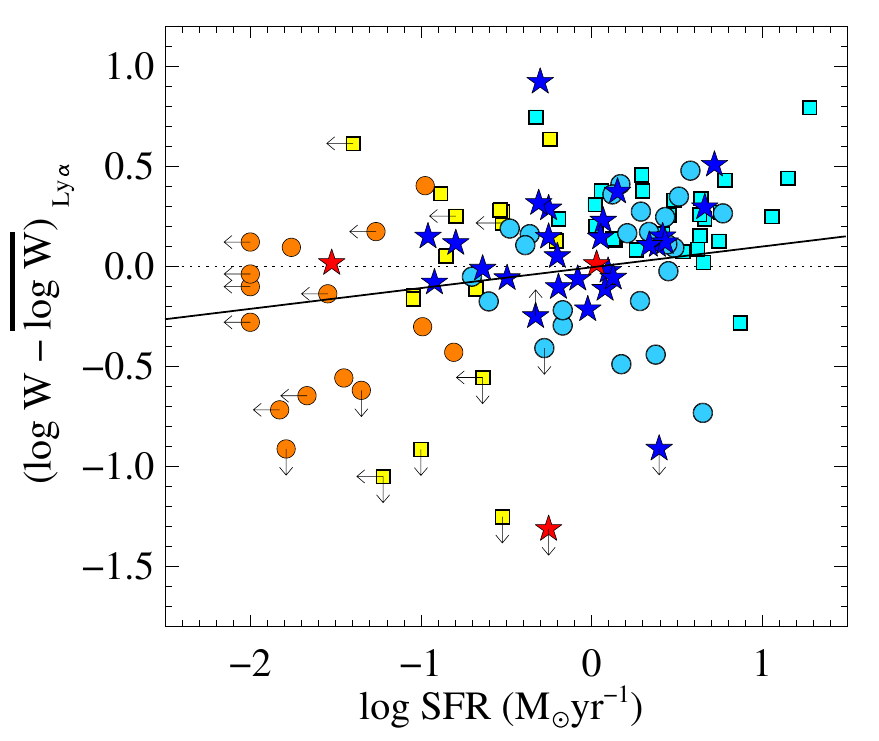}
\hspace{0.5cm}
\includegraphics[trim = 0mm 114mm 00mm 40mm, clip, angle=-0, height=2.5in]{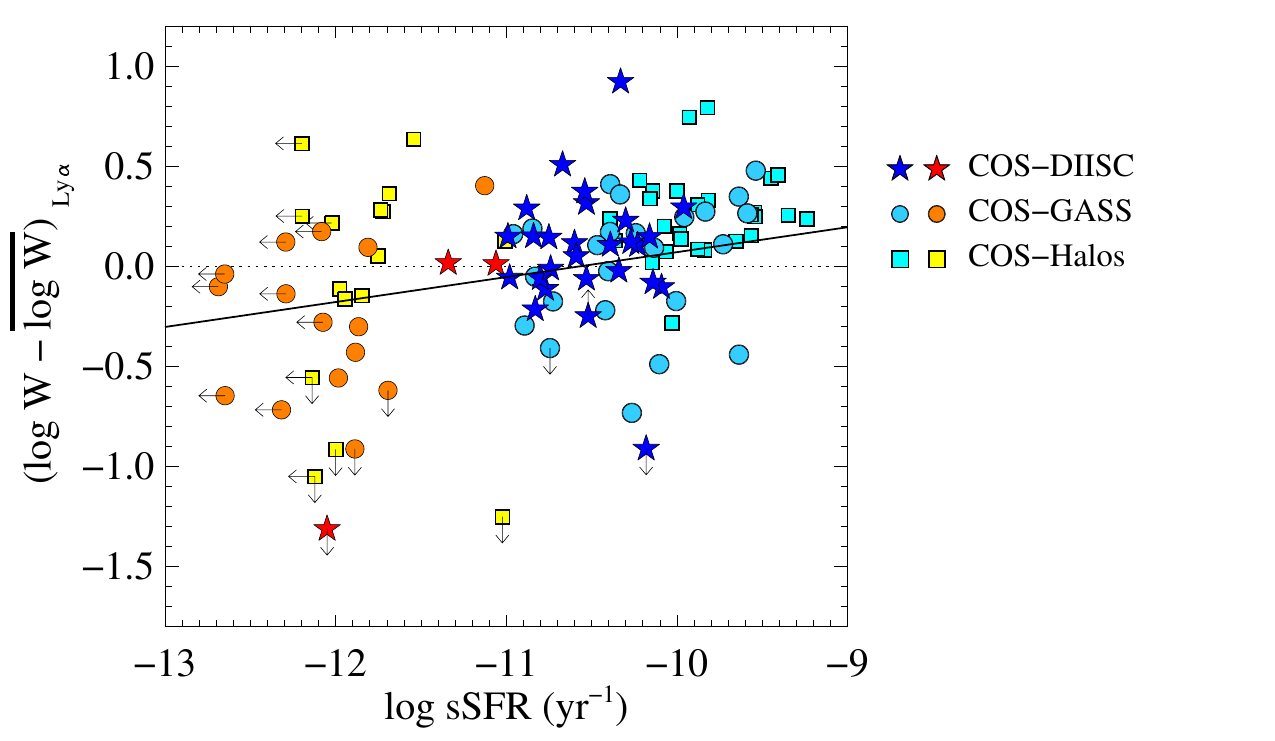}

\caption{Excess in the \Lya equivalent width, $\rm log~W_{Ly\alpha}^{excess} = log~W_{Ly\alpha} - \overline{log W_{Ly\alpha} }$, as a function of \HI\ mass (top left), gas fraction (top right), SFR (bottom left), and sSFR (bottom right). The color of the symbols represents their sSFR with blue, light blue, and cyan for galaxies with log~sSFR $>-11$ and red, orange, and yellow for galaxies with log~sSFR $<-11$. The shapes of the symbols are indicated in the legends. The best-fit lines to the data are shown as the black solid line, and the parameters (slope and intercept) for the lines are presented in Table~\ref{tab:Wexcess_asur_fits}.}
 \label{Wexcess_MHI_SFR}
\end{figure*}

Figure~\ref{Wexcess_MHI_SFR} shows the distribution of the excess \Lya equivalent width as a function of \HI\ mass, gas fraction $\rm M_{HI}/(M_{\star}+M_{HI})$, SFR, and sSFR. We found a significant positive correlation between the equivalent width of \Lya with \HI\ mass and \HI\ gas fraction at a confidence level of 99.8\% and 99.9\%, respectively. The strengths were estimated using the survival analysis software package ASURV \citep{asurv} using a combined sample from the COS-DIISC and COS-GASS surveys. \HI\ masses of the COS-Halos sample are not available and hence are not included in the analysis. This correlation is not surprising as we already discovered a much stronger correlation (and lower scatter) between the  \Lya equivalent width and $\rho/R_{HI}$. The slope and intercept for the best-fit lines are presented in Table~\ref{tab:Wexcess_asur_fits}.

Similarly, SFR and sSFR also show a strong correlation with the excess \Lya equivalent width at a confidence level of 99.99\% or higher. For this analysis, we combined data from COS-DIISC, COS-GASS, and COS-Halos surveys, thereby creating a sample 50\% bigger than the COS-DIISC and COS-GASS combined. Lower panels of Figure~\ref{Wexcess_MHI_SFR} show the data and the best-fit lines. We estimate a standard deviation of 0.4~dex between the data and the best-fit lines for both plots. 

The offsets from the best-fit line relating $W_{Ly\alpha}^{excess}$ and one of the four parameters -- $\rm M_{HI}$, $\rm M_{HI}/M_{total}$, SFR, sSFR -- do not show dependence on one of the other parameters. For example, the offset from the best-fit line relating \HI mass and gas fraction to $W_{Ly\alpha}^{excess}$ (top panels of Figure~\ref{Wexcess_MHI_SFR}) show a balanced distribution of red and blue galaxies on either side of the best-fit lines. This is not surprising as the gas mass and gas fractions are known to correlate with the SFR and sSFR. Therefore, the source of the divergence could either be due to stochasticities or an independent parameter not considered here. We refrain from fitting further parameters as we do not see any improvement in the dispersion. Therefore, we conclude that a combination of $\rho/R_{vir}$ and one of the four parameters discussed here is better than $\rho/R_{vir}$ alone in describing the equivalent width. However, addition parameterization is not beneficial.

\subsection{CGM gas profile in terms of $\rho/\rm R_{vir}$ and star-formation rates  } \label{sec:discussion_Wexcess}

\begin{figure*}[t]
\includegraphics[trim = 0mm 114mm 67mm 40mm, clip, angle=-0, height=2.5in]{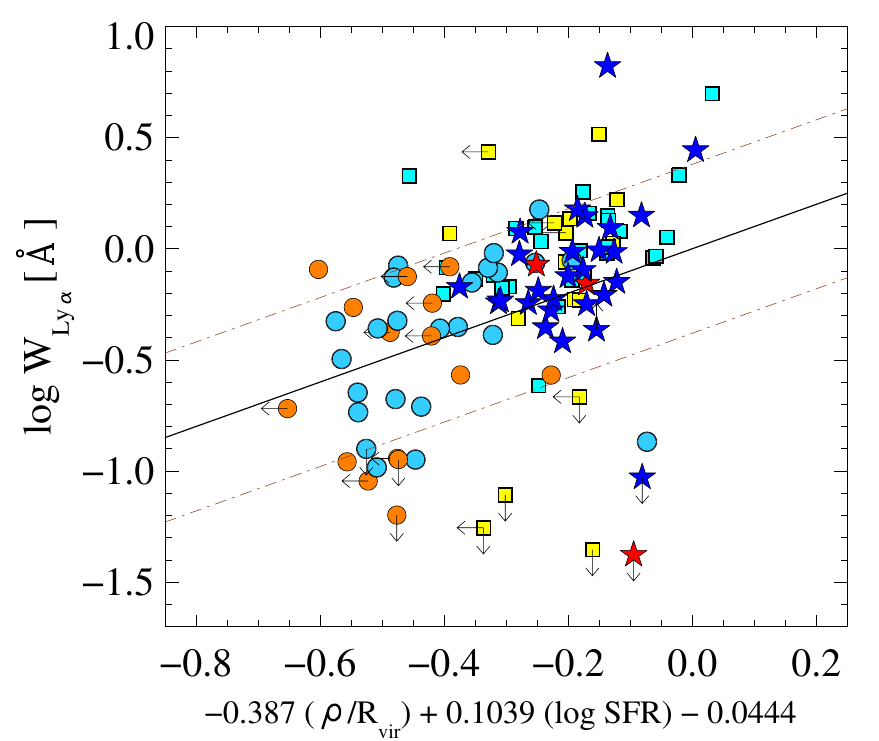}
\includegraphics[trim = 0mm 114mm 00mm 40mm, clip, angle=-0, height=2.5in]{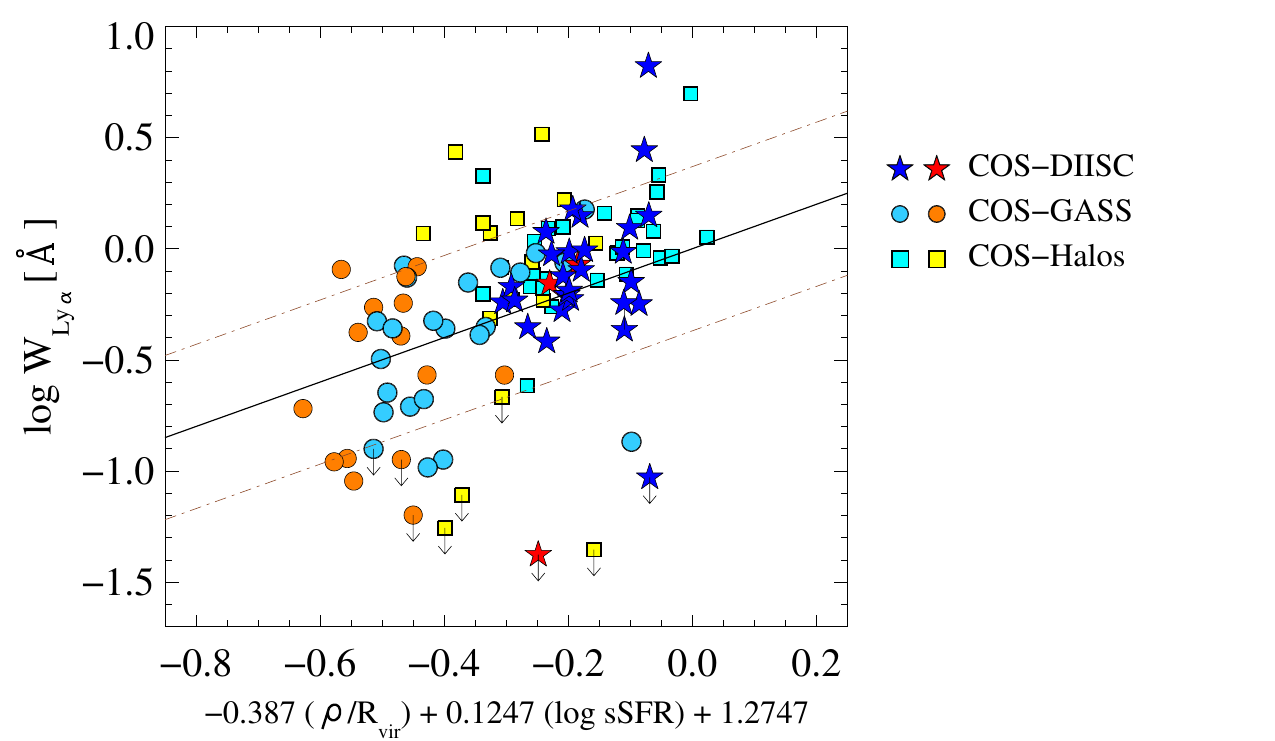}
\caption{\Lya equivalent width as a function of SFR and sSFR corrected normalized impact parameter, $\rho/\rm R_{vir}$, based on the relationship described in equations~\ref{Eq_W_Rvir_SFR} and \ref{Eq_W_Rvir_sSFR}.  The solid line shows a perfect match with the best-fit line, i.e., $\rm log~W_{Ly\alpha} =\overline{log W_{Ly\alpha}}$, and the two brown dot-dashed lines indicate the dispersion ($\pm 1~\sigma$) in the points. 25 (30) of the 113 points show offsets larger than one standard deviation in SFR (sSFR).}
 \label{W_parameterization}
\end{figure*}

\begin{figure*}[t]
\includegraphics[trim = 10mm 112mm 0mm 20mm, clip, angle=-0, height=2.95in]{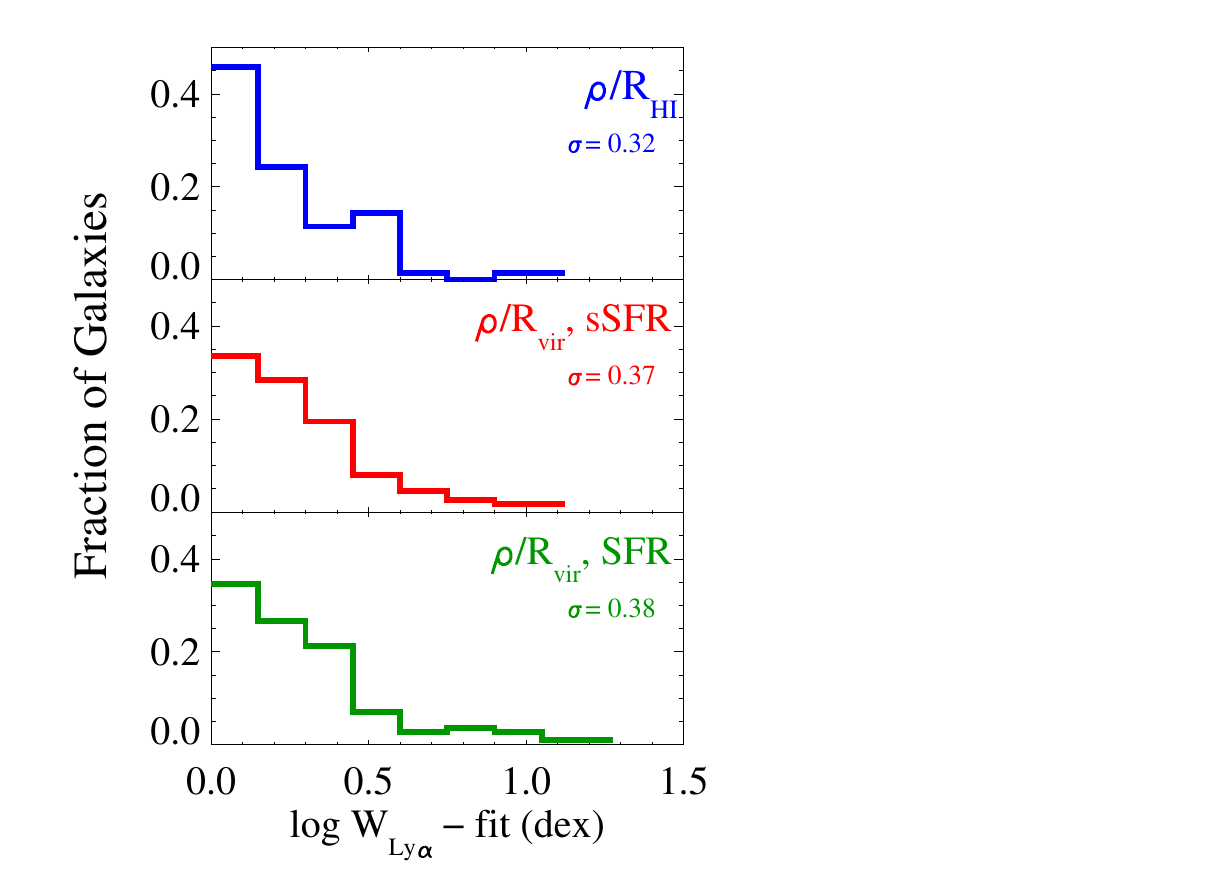}
\hspace{-3cm}
\caption{A comparison of the distribution of the data with respect to the best-fit lines describing the relationship between \Lya equivalent width and $\rho/\rm R_{HI}$ (Eq.~\ref{Eq_rhoRHI}),   $\rho/\rm R_{vir}$ \& sSFR (Eq.~\ref{Eq_W_Rvir_sSFR}) and $\rho/\rm R_{vir}$ \& SFR (Eq.~\ref{Eq_W_Rvir_SFR}). We find that the fraction of data points with 0.15~dex of the fit also drops from 46\% to 35\% to 34\% for the fit for $\rho/R_{HI}$, $\rho/\rm R_{vir}$ \& SFR, and $\rho/\rm R_{vir}$ \& sSFR, respectively. We also find that the dispersion is smallest for $\rho/R_{HI}$ at 0.32 dex followed by $\rho/\rm R_{vir}$ \& sSFR and $\rho/\rm R_{vir}$ \& SFR  at 0.37 and 0.38~dex respectively. These distributions indicate that $\rho/\rm R_{HI}$ (Eq.~\ref{Eq_rhoRHI}) is still more robust at predicting the \Lya equivalent width compared to $\rho/\rm R_{vir}$ even after accounting for star-formation activity (SFR or sSFR).} 
 \label{Histogram_2}
\end{figure*}

\HI\ measurement for higher redshift galaxies (even at $z=0.1$ and beyond) are not readily available (except for a small fraction of the sky \citep{Hess19} or stacks \citep[e.g.][]{Chowdhury24, Sinigaglia24}), nor are they easy to acquire. The most commonly available information on the foreground galaxy is its position, redshift, stellar mass, and SFR. Through prescriptions from theoretical models, one can derive a halo mass from stellar mass and, in turn, a virial radius. In light of this, there is value to exploring if the star formation within galaxies may allow us to explain the large scatter in the relationship between \Lya equivalent width and virial radius normalized impact parameter ($\rm \rho/R_{vir}$). Our goal is to parameterize the relationship by adding SFR or sSFR to find a tighter relationship. We use a combination of impact parameter, virial radius, and SFR or sSFR to minimize the scatter and relate the observed \Lya equivalent width to those parameters using the following relations as derived from the best fit to the combined data from COS-DIISC, COS-Halos, and COS-GASS. 

\vspace{-0.3cm}
\begin{equation}\label{Eq_W_Rvir_SFR}
\rm log~W_{Ly\alpha} = - 0.387~ (\rho/ \rm R_{vir}) + 0.1039~log~SFR - 0.0444
\end{equation} 
\begin{equation} \label{Eq_W_Rvir_sSFR}
\rm log~W_{Ly\alpha} = - 0.387~ (\rho/ \rm R_{vir}) + 0.1247~log~sSFR + 1.2747
\end{equation}

Figure~\ref{W_parameterization} shows the data in terms of these equations. The solid line represents a perfect match with the best-fit line, i.e., $\rm log~W_{Ly\alpha} =\overline{log W_{Ly\alpha} }$. The scatter in the two relationships is 0.38 and 0.37~dex, respectively, which is still higher than the relationship between $\rm log~W_{Ly\alpha}$ and $\rm log~\rho/R_{HI}$ described in Eq~\ref{Eq_rhoRHI}. This indicates that combined stellar mass and SFR are not as good indicators of the gas profile in the CGM as the size of the gas disk. The same data are also shown in Figure~\ref{Histogram_2} as offset from the best-fit lines described by Eqs.~\ref{Eq_rhoRHI}, \ref{Eq_W_Rvir_SFR}, \& \ref{Eq_W_Rvir_sSFR}. The fractional distribution of galaxies with the lowest offset from the line is seen for  $\rho/R_{HI}$ (Eqs.~\ref{Eq_rhoRHI}) followed by the combination of $\rho/R_{vir}$ and sSFR (Eqs.~\ref{Eq_W_Rvir_sSFR}), and $\rho/R_{vir}$ and SFR (Eqs.~\ref{Eq_W_Rvir_SFR}).

This result reiterates the idea that the CGM properties are more closely related to proximity from the gas disk rather than the center from the galaxy ($\rho$), position within the halo ($\rho/R_{vir}$), stellar mass ($M_{\star}$), or star formation rates (SFR, sSFR).
In the context of the baryon cycle, we hypothesize that this result is a consequence of the process that leads to gas accretion into the disk. Each step in the baryon cycle -- condensation of gas from the CGM into the \HI\ disks, conversion of \HI into molecular gas, transformation of molecular gas into stars, and enrichment of the CGM with outflowing material from young star-forming regions -- takes tens to hundreds of millions of years. Therefore, although the processes are connected, the correlation is expected to weaken when we compare processes that are steps apart in the baryon cycle. The strongest relationship between the neutral gas content of the CGM and the \HI radii suggests that neutral gas in the CGM is predominantly associated with the accretion process that builds the disk. In addition, such a strong correlation can only be supported if gas accretion into the disk is continuous and symmetric as opposed to sporadic or directional. In the cosmological context, this view of the disk buildup is supported by the fact that galaxies with larger \HI disks (and masses) have a stronger incidence of \Lya absorbers beyond their halos \citep{borthakur22, RyanWeber06}, thus confirming the presence of ample material to support long-term continuous accretion.

Extending this idea further, we also predict that a good fraction of the CGM material detected in these sightlines should show co-rotating kinematics with the \HI disk. Again, this need not be gas in a thin disk but may take the other volume-filling shapes that may eventually settle in the disk. An extended morphology would make the detection rates and strengths independent of the orientation of the sightline, which is what we observed. This does not contradict orientation dependencies found by other studies \citep{Bordoloi14, Nielsen15, pointon19, Peroux20} as we are looking very close to the gas disk, and the projection effect from disk morphology (thin, thick, or flared) will be significant, which may not be the case at slightly larger distances.

\section{Summary}\label{sec:summary}

We presented the rationale, design, and observations for the Deciphering the Interplay Between the Interstellar medium, Stars, and the Circumgalactic Medium (DIISC) Survey. The survey aims to characterize the correlations between the properties of the CGM, ISM, and stars to identify the flow of gas into, within, and out of galaxies. Our sample consisted of 33 galaxies with UV-bright QSO sightlines within $\sim$4 times the \HI radius. We acquired QSO absorption spectroscopy with the COS abroad HST, \HI 21cm and radio continuum imaging with the VLA, optical broadband and H$\alpha$ imaging with the VATT and LBT, and multi-slit spectroscopy with Binoscpec on the MMT.  
We have produced resolved images of these galaxies in \HI 21cm, H$\alpha$, and broadband optical and UV photometry, as well as partially resolved maps of dust content and metallicity. Future (and past) papers will focus on studies connecting the properties of the circumgalactic gas to the baryonic phases traced by gas and stars in the disk \citep{gim21, padave21, padave24}.

In this paper, we presented one of the first results from the COS-DIISC survey. We found that the equivalent width of \Lya absorption is strongly correlated (inverse correlation) to the proximity of the QSO sightline from the \HI disk, i.e., $\rho/R_{HI}$. The correlation was seen at a confidence level of 99.99\%. $\rho/R_{HI}$ is a better indicator of the equivalent width of \Lya absorption in the CGM than traditionally used parameters like virial radius normalized impact parameter, i.e., $\rho/R_{vir}$. We also investigated if the addition of SFR and sSFR to the $\rho/R_{vir}$ relationship could reduce the scatter and provide a better fit than $\rho/R_{HI}$. We found the fit described by Eq.~\ref{Eq_rhoRHI} relating the equivalent width of \Lya and $\rho/R_{HI}$ to have the lowest dispersion among all the parameters explored, including  $\rho/R_{vir}$, $\rm \rho/M_{vir}$, SFR, sSFR.

The strong anti-correlation and tighter data fit indicate that the CGM in the disk-CGM interface is likely physically connected to the disk through processes that build the disk. 
Even though $\rm R_{HI}$ and $\rm M_{HI}$ are degenerate, we conclude that the size of the gas disk (which is well characterized by $\rm R_{HI}$) is more likely to be connected to CGM properties.
The large timescale required for gas to move and transition between different phases seen in the baryon cycle suggests that the strongest correlation is observed for phases directly connected and adjacent in the baryon cycle. Therefore, the correlation is likely representative of the gas inflow process.

If our hypothesis is accurate, we should see a corotating disk extending far beyond the optical and \HI 21cm disk in galaxies. Further investigation in the context of CGM-disk kinematics and comparisons to prediction from simulations will provide a clearer picture of the processes that lead to disk buildup in the local galaxies.

%\begin{acknowledgments}

%We thank the referee for their constructive comments.
~\\
{\em Acknowledgements} 

SB, TM, and MP thank the funding support from STScI through HST GO grant 14071.
SB, HG, MP, AO, and BK acknowledge funding support through NSF AAG grants 2108159 and 2009409.
SB, MP acknowledges funding support through NASA ADAP grant 80NSSC21K0643.
DN acknowledges funding from the Deutsche Forschungsgemeinschaft (DFG) through an Emmy Noether Research Group (grant number NE 2441/1-1).

We thank the staff and support scientists at Space Telescope Science Institute, Array Operations Center of the National Radio Astronomy Observatory, Steward Observatory (LBT, MMT, Bok telescope, VATT), Smithsonian Center of Astrophysics, and Arizona Radio Observatory for their help and support in carrying the observations.
We thank  Hsiao-Wen Chen, Xavier Prochaska, Rongmon Bordoloi, Todd Tripp, and Jessica Werk for the discussions over the years that positively influenced this work. 
SB thanks Anika B. Srivastava for her unique suggestions for improving the presentation of the figures. 

SB also acknowledges the Indigenous peoples, including the Akimel O'odham (Pima) and Pee Posh (Maricopa) Indian Communities, whose care and keeping of the land has enabled her to be at ASU's Tempe campus in the Salt River Valley, where this work was conducted. 

%\end{acknowledgments}

\vspace{5mm}
\facilities{HST(COS), VLA, GALEX, VATT, LBT, MMT(Binospec), WISE, Sloan, Arecibo }

\bibliographystyle{apj}	       
\bibliography{DIISC_survey}

%% File created on Thu Jul 18 23:56:45 2024 with code /Users/sanch/Dropbox/COS-Disk/Sample/Create_table1.pro
\begin{longrotatetable}
\begin{deluxetable*}{ccc rrr rrc ccr c}
\tablecaption{DIISC galaxies and their derived properties\label{tbl-samp}}
\tablewidth{0pt}
%\rotate
%\tabletypesize{\small}
\tablehead{
\colhead{\#} &\colhead{Galaxy}& \colhead{Name}& \colhead{R.A.} & \colhead{Decl.} & \colhead{Distance}  & \colhead{\mstar}  &   \colhead{$\rm M_{HI} $} & \colhead{$\rm R_{HI}$} & \colhead{$\log_{10}$ SFR} & \colhead{b/a} & \colhead{P.A.} &\colhead{Inclination} \\
\colhead{}  &   \colhead{}  & \colhead{} &   \colhead{}  & \colhead{}  & \colhead{(Mpc)}  &  \colhead{(log $M_\odot$)}    &   \colhead{(log $M_\odot$)} & \colhead{(kpc)} & \colhead{($M_\odot$~yr$^{-1}$)} &   \colhead{}&  \colhead{($^{\circ}$)} &\colhead{($^{\circ}$)}  \\ }
\decimalcolnumbers
\startdata
       1 & J002400+154623 & NGC99 & 5.997 & 15.7704 & 79.4$^a$ & 10.62 & 10.35 & 52.0 & 0.39 & 0.94 & 92 & 20 \\
       2 & J023526--092155 & NGC988 & 38.866 & -9.3563 & 19.0$^b$ & 9.82 & 9.33 & 14.7 & -1.52 & 0.37 & 114 & 71 \\
       3 & J083221+243142 & KUG0829+246B & 128.090 & 24.5285 & 183.4$^b$ & 10.37 & 10.00 & 33.4 & 0.07 & 0.75 & 100 & 42 \\
       4 & J083531+250132 & NGC2611 & 128.872 & 25.0275 & 77.3$^b$ & 10.62 & 9.59 & 20.2 & 0.66 & 0.38 & 52 & 71 \\
       5 & J091729+272039 & J0917+2720 & 139.377 & 27.3441 & 199.9$^b$ & 10.63 & 9.82 & 26.8 & -0.25 & 0.89 & 78 & 27 \\
       6 & J102425+242425 & J1024+2424 & 156.108 & 24.4078 & 89.6$^b$ & 9.43 & 9.66 & 22.0 & -0.80 & 0.30 & 178 & 77 \\
       7 & J104330+245514 & NGC3344 & 160.880 & 24.9225 & 8.3$^a$ & 10.19 & 9.44 & 16.8 & -0.33 & 1.00 & 155 & 6 \\
       8 & J104357+114215 & M95/NGC3351 & 160.990 & 11.7038 & 9.3$^a$ & 10.90 & 9.00 & 9.7 & -0.10 & 0.71 & 7 & 46 \\
       9 & J105203+100852 & NGC3433 & 163.016 & 10.1483 & 44.6$^b$ & 10.85 & 10.16 & 41.0 & 0.08 & 0.93 & 46 & 23 \\
      10 & J105951+251700 & 2MASXJ1059+2516 & 164.969 & 25.2760 & 85.5$^b$ & 9.90 & 9.33 & 14.6 & -0.19 & 0.65 & 67 & 51 \\
      11 & J110002+144959 & NGC3485 & 165.010 & 14.8416 & 29.4$^b$ & 10.45 & 9.75 & 24.7 & -0.08 & 0.82 & 89 & 36 \\
      12 & J112426+112034 & NGC3666 & 171.109 & 11.3422 & 17.1$^a$ & 10.31 & 9.44 & 16.9 & -0.49 & 0.26 & 96 & 80 \\
      13 & J113316+242645 & NGC3728 & 173.316 & 24.4469 & 101.3$^b$ & 11.26 & 10.58 & 68.8 & 0.42 & 0.60 & 21 & 55 \\
      14 & J113458+255605 & AGC723761 & 173.739 & 25.9456 & 141.7$^b$ & 10.03 & 9.60 & 20.5 & -0.96 & 0.59 & 29 & 55 \\
      15 & J114058+112822 & NGC3810 & 175.245 & 11.4711 & 15.3$^a$ & 10.44 & 9.43 & 16.7 & 0.10 & 0.69 & 27 & 48 \\
      16 & J114430+070328 & J1144+0703 & 176.127 & 7.0604 & 218.7$^b$ & 11.09 & 10.29 & 47.9 & 0.03 & 0.93 & 24 & 23 \\
      17 & J115704+090552 & AGC213643 & 179.259 & 9.1028 & 160.7$^b$ & 9.80 & 9.99 & 33.3 & -0.80 & 0.32 & 120 & 75 \\
      18 & J115718+113907 & 2MASXJ1157+1139 & 179.325 & 11.6559 & 89.2$^b$ & 9.91 & 9.54 & 19.0 & -0.25 & 0.45 & 74 & 66 \\
      19 & J120929+261331 & UGC7138 & 182.370 & 26.2263 & 34.6$^b$ & 9.22 & 9.11 & 11.2 & -0.92 & 0.63 & 89 & 53 \\
      20 & J122154+042837 & M61/NGC4303 & 185.479 & 4.4735 & 18.7$^a$ & 11.39 & 9.88 & 28.9 & 0.72 & 0.93 & 138 & 23 \\
      21 & J122255+154929 & M100/NGC4321 & 185.729 & 15.8221 & 13.9$^a$ & 10.58 & 9.38 & 15.6 & 0.40 & 0.91 & 80 & 25 \\
      22 & J123106+120150 & IC3440 & 187.772 & 12.0299 & 120.5$^b$ & 10.22 & 9.73 & 24.2 & -0.31 & 0.78 & 81 & 40 \\
      23 & J130130+275232 & NGC4921 & 195.359 & 27.8859 & 88.3$^a$ & 11.80 & 9.20 & 12.6 & -0.25 & 0.93 & 177 & 23 \\
      24 & J130330+263235 & UGC8161 & 195.871 & 26.5505 & 90.2$^b$ & 10.81 & 9.92 & 30.4 & -0.02 & 0.37 & 144 & 71 \\
      25 & J131443+260516 & J1314+2605 & 198.671 & 26.0875 & 184.6$^b$ & 10.70 & 9.99 & 33.1 & 0.43 & 0.69 & 130 & 47 \\
      26 & J135428+144030 & KUG1352+149 & 208.625 & 14.6757 & 177.0$^b$ & 10.69 & 10.27 & 47.1 & 0.15 & 0.32 & 123 & 75 \\
      27 & J141512+044953 & UGC9120 & 213.801 & 4.8243 & 77.3$^b$ & 11.11 & 9.66 & 22.0 & 0.13 & 0.81 & 14 & 37 \\
      28 & J143222+095551 & KUG1429+101 & 218.087 & 9.9335 & 18.5$^b$ & 8.89 & 8.59 & 5.9 & -2.00 & 0.88 & 114 & 29 \\
      29 & J143245+095343 & NGC5669 & 218.183 & 9.8917 & 14.5$^a$ & 9.52 & 9.23 & 13.0 & -0.54 & 0.67 & 58 & 49 \\
      30 & J152447+042057 & 2MASXJ1524+0421 & 231.205 & 4.3603 & 158.5$^b$ & 10.73 & 9.78 & 25.5 & 0.34 & 0.56 & 15 & 58 \\
      31 & J153313+150147 & LEDA140527 & 233.307 & 15.0234 & 179.8$^b$ & 10.03 & 9.76 & 24.9 & -0.30 & 0.81 & 174 & 37 \\
      32 & J153343+150042 & NGC5951 & 233.429 & 15.0072 & 27.1$^a$ & 10.10 & 9.53 & 18.7 & -0.64 & 0.23 & 4 & 83 \\
      33 & J155807+120411 & IC1149 & 239.533 & 12.0703 & 62.8$^b$ & 10.80 & 9.70 & 23.3 & 0.05 & 0.83 & 172 & 35 \\
      34 & J155850+125330 & KUG1556+130 & 239.708 & 12.8920 & 136.9$^b$ & 10.39 & 9.92 & 30.5 & -0.20 & 0.41 & 111 & 68 \\
\hline 
\enddata
\tablecomments{$^a$ Distance via direct method;  $^b$ Distance via the flow model}
\end{deluxetable*}
\end{longrotatetable}

%% File created on Mon Feb 12 15:47:37 2024 with code /Users/sanch/Dropbox/COS-Disk/Sample/Create_table2.pro
\begin{longrotatetable}
\begin{deluxetable*}{ccc rcc rrr rrr rrr}
\tablecaption{QSO sightlines to DIISC galaxies \label{tbl-samp_qso}}
\tablewidth{0pt}
%\rotate
%\tabletypesize{\small}
\tablehead{
\colhead{\#} & \colhead{Galaxy}  & \colhead{$\rm Z_{gal}$} & \colhead{ \mstar}  &   \colhead{$\rm M_{halo}$}      & \colhead{$\rm R_{vir}$} & \colhead{$\rm R_{HI}$} & \colhead{QSO} & \colhead{R.A.} & \colhead{Decl.} &  \colhead{$\rm Z_{QSO}$}  & \colhead{$\rho$} & \colhead{$\rm \rho/R_{HI}$} & \colhead{$\rm \rho/R_{vir}$} & \colhead{$\rm \Theta $}  \\
\colhead{}     & \colhead{}     &   \colhead{}  &  \colhead{($\rm log~M_\odot$)} &  \colhead{($\rm log~M_\odot$)}&     \colhead{(kpc)}     &     \colhead{(kpc)}    & \colhead{}    & \colhead{}     & \colhead{}      &  \colhead{}               & \colhead{(kpc)}  &                             &                              &\colhead{($^{\circ}$)}  \\ }
\decimalcolnumbers
\startdata
       1 & J002400+154623 & 0.0177 & 10.62 & 11.7 & 207 & 52 & J002330+154745 & 5.878 & 15.7960 & 0.41 & 163 &3.1 &0.8 &$\rm ~ ~ 0^a$ \\
       2 & J023526--092155 & 0.0050 & 9.82 & 11.5 & 181 & 15 & J023529--092511 & 38.870 & -9.4200 & 1.78 & 21 &1.4 &0.1 &62~ \\
       3 & J083221+243142 & 0.0432 & 10.37 & 11.5 & 173 & 33 & J083220+243101 & 128.084 & 24.5170 & 1.30 & 41 &1.2 &0.2 &74~ \\
       4 & J083531+250132 & 0.0175 & 10.62 & 11.7 & 207 & 20 & J083535+245941 & 128.899 & 24.9950 & 0.33 & 55 &2.7 &0.3 &89~ \\
       5 & J091729+272039 & 0.0469 & 10.63 & 11.7 & 203 & 27 & J091728+271951 & 139.369 & 27.3310 & 0.08 & 52 &1.9 &0.3 &$\rm ~ ~ 0^a$ \\
       6 & J102425+242425 & 0.0209 & 9.43 & 11.1 & 124 & 22 & J102416+242212 & 156.070 & 24.3700 & 0.77 & 80 &3.6 &0.6 &45~ \\
       7 & J104330+245514 & 0.0020 & 10.19 & 11.4 & 165 & 17 & J104241+250122 & 160.672 & 25.0230 & 0.34 & 31 &1.8 &0.2 &$\rm ~ ~ 0^a$ \\
       8 & J104357+114215 & 0.0026 & 10.90 & 12.3 & 334 & 10 & J104335+115129 & 160.900 & 11.8580 & 0.79 & 29 &3.0 &0.1 &37~ \\
       9 & J105203+100852 & 0.0091 & 10.85 & 12.0 & 252 & 41 & J105220+101754 & 163.086 & 10.2980 & 0.25 & 128 &3.1 &0.5 &$\rm ~ ~ 0^a$ \\
      10 & J105951+251700 & 0.0207 & 9.90 & 11.3 & 145 & 15 & J105958+251709 & 164.995 & 25.2860 & 0.66 & 38 &2.6 &0.3 &0~ \\
      11 & J110002+144959 & 0.0048 & 10.45 & 11.6 & 188 & 25 & J105945+144143 & 164.938 & 14.6950 & 0.63 & 83 &3.4 &0.4 &$\rm ~ ~ 0^a$ \\
      12 & J112426+112034 & 0.0035 & 10.31 & 11.5 & 174 & 17 & J112439+113117 & 171.165 & 11.5210 & 0.14 & 56 &3.3 &0.3 &79~ \\
      13 & J113316+242645 & 0.0232 & 11.26 & 12.5 & 382 & 69 & J113325+242328 & 173.358 & 24.3910 & 0.53 & 120 &1.7 &0.3 &55~ \\
      14 & J113458+255605 & 0.0321 & 10.03 & 11.3 & 150 & 21 & J113457+255528 & 173.740 & 25.9250 & 0.71 & 51 &2.5 &0.3 &30~ \\
      15 & J114058+112822 & 0.0033 & 10.44 & 11.6 & 187 & 17 & J114046+113650 & 175.192 & 11.6140 & 0.69 & 41 &2.4 &0.2 &47~ \\
      16 & J114430+070328 & 0.0507 & 11.09 & 12.6 & 385 & 48 & J114434+070516 & 176.145 & 7.0880 & 0.07 & 125 &2.6 &0.3 &$\rm ~ ~ 0^a$ \\
      17 & J115704+090552 & 0.0367 & 9.80 & 11.2 & 139 & 33 & J115709+090607 & 179.291 & 9.1020 & 0.30 & 88 &2.6 &0.6 &29~ \\
      18 & J115718+113907 & 0.0213 & 9.91 & 11.3 & 146 & 19 & J115722+114041 & 179.344 & 11.6780 & 0.29 & 45 &2.4 &0.3 &34~ \\
      19 & J120929+261331 & 0.0072 & 9.22 & 11.0 & 119 & 11 & J120917+261612 & 182.322 & 26.2700 & 0.58 & 37 &3.3 &0.3 &46~ \\
      20 & J122154+042837 & 0.0052 & 11.39 & 12.7 & 457 & 29 & J122138+043026 & 185.408 & 4.5070 & 0.09 & 26 &0.9 &0.1 &$\rm ~ ~ 0^a$ \\
      21 & J122255+154929 & 0.0053 & 10.58 & 11.7 & 204 & 16 & J122330+154507 & 185.879 & 15.7520 & 0.08 & 39 &2.5 &0.2 &$\rm ~ ~ 0^a$ \\
      22 & J123106+120150 & 0.0260 & 10.22 & 11.4 & 163 & 24 & J123113+120307 & 187.805 & 12.0520 & 0.12 & 83 &3.4 &0.5 &25~ \\
      23 & J130130+275232 & 0.0182 & 11.80 & 13.8 & 1009 & 13 & J130128+275106 & 195.367 & 27.8520 & 0.24 & 53 &4.3 &0.1 &$\rm ~ ~ 0^a$ \\
      24 & J130330+263235 & 0.0223 & 10.81 & 11.9 & 240 & 30 & J130345+263314 & 195.941 & 26.5540 & 0.44 & 99 &3.2 &0.4 &58~ \\
      25 & J131443+260516 & 0.0431 & 10.70 & 11.8 & 215 & 33 & J131447+260624 & 198.696 & 26.1070 & 0.07 & 96 &2.9 &0.4 &81~ \\
      26 & J135428+144030 & 0.0418 & 10.69 & 11.8 & 213 & 47 & J135426+144151 & 208.609 & 14.6980 & 0.21 & 84 &1.8 &0.4 &21~ \\
      27 & J141512+044953 & 0.0192 & 11.11 & 12.3 & 323 & 22 & J141505+044547 & 213.775 & 4.7630 & 0.24 & 90 &4.1 &0.3 &$\rm ~ ~ 0^a$ \\
      28 & J143222+095551 & 0.0046 & 8.89 & 10.8 & 107 & 6 & J143216+095520 & 218.070 & 9.9220 & 0.77 & 6 &1.1 &0.1 &$\rm ~ ~ 0^a$ \\
      29 & J143245+095343 & 0.0046 & 9.52 & 11.1 & 130 & 13 & J143216+095520 & 218.070 & 9.9220 & 0.77 & 29 &2.2 &0.2 &47~ \\
      30 & J152447+042057 & 0.0374 & 10.73 & 11.8 & 221 & 26 & J152447+041920 & 231.199 & 4.3220 & 0.71 & 107 &4.2 &0.5 &6~ \\
      31 & J153313+150147 & 0.0429 & 10.03 & 11.3 & 149 & 25 & J153314+150102 & 233.310 & 15.0170 & 0.09 & 22 &0.9 &0.1 &$\rm ~ ~ 0^a$ \\
      32 & J153343+150042 & 0.0059 & 10.10 & 11.4 & 158 & 19 & J153314+150102 & 233.310 & 15.0170 & 0.09 & 55 &2.9 &0.3 &89~ \\
      33 & J155807+120411 & 0.0156 & 10.80 & 11.9 & 239 & 23 & J155821+120533 & 239.591 & 12.0930 & 0.57 & 67 &2.9 &0.3 &$\rm ~ ~ 0^a$ \\
      34 & J155850+125330 & 0.0347 & 10.39 & 11.5 & 176 & 31 & J155855+125555 & 239.731 & 12.9320 & 0.29 & 109 &3.6 &0.6 &82~ \\
\hline 
\enddata
\end{deluxetable*}
\end{longrotatetable}

%% File created on Tue Dec 20 12:30:39 2022 with code /Users/sanch/Dropbox/COS-Disk/Analysis/Plots/Create_table3.pro

\begin{deluxetable*}{ccc rcc ccc rrr ccccc}
\tablecaption{Lyman~$\alpha$ equivalent width in the CGM of DIISC galaxies \label{tbl-W_LymanA}}
\tablewidth{0pt}
%\rotate
%\tabletypesize{\small}
\tablehead{
\colhead{\#} & \colhead{Galaxy}  & \colhead{$\rm Z_{gal}$} & \colhead{QSO}  & \colhead{$\rho$} & \colhead{$\rm \rho/R_{HI}$} & \colhead{$\rm W_{Ly\alpha}$} \\
\colhead{}   & \colhead{}        &   \colhead{}            &   \colhead{}   & \colhead{(kpc)}  & \colhead{}                  & \colhead{($\rm m\AA$)}       \\ }
\startdata
       1 & J002400+154623 & 0.0177 & J002330+154745 & 163 &3.1&574$\pm$101 \\
       2 & J023526--092155 & 0.0050 & J023529--092511 & 21 &1.4&848$\pm$228 \\
       3 & J083221+243142 & 0.0432 & J083220+243101 & 41 &1.2&1242$\pm$29 \\
       4 & J083531+250132 & 0.0175 & J083535+245941 & 55 &2.7&1411$\pm$25 \\
       5 & J091729+272039 & 0.0469 & J091728+271951 & 52 &1.9&1405$\pm$41 \\
       6 & J102425+242425 & 0.0209 & J102416+242212 & 80 &3.6&$-$ \\
       7 & J104330+245514 & 0.0020 & J104241+250122 & 31 &1.8&$>$~432 \\
       8 & J104357+114215 & 0.0026 & J104335+115129 & 29 &3.0&1746$\pm$22 \\
       9 & J105203+100852 & 0.0091 & J105220+101754 & 128 &3.1&444$\pm$39 \\
      10 & J105951+251700 & 0.0207 & J105958+251709 & 38 &2.6&563$\pm$22 \\
      11 & J110002+144959 & 0.0048 & J105945+144143 & 83 &3.4&529$\pm$22 \\
      12 & J112426+112034 & 0.0035 & J112439+113117 & 56 &3.3&594$\pm$28 \\
      13 & J113316+242645 & 0.0232 & J113325+242328 & 120 &1.7&967$\pm$58 \\
      14 & J113458+255605 & 0.0321 & J113457+255528 & 51 &2.5&944$\pm$30 \\
      15 & J114058+112822 & 0.0033 & J114046+113650 & 41 &2.4&709$\pm$42 \\
      16 & J114430+070328 & 0.0507 & J114434+070516 & 125 &2.6&697$\pm$18 \\
      17 & J115704+090552 & 0.0367 & J115709+090607 & 88 &2.6&674$\pm$22 \\
      18 & J115718+113907 & 0.0213 & J115722+114041 & 45 &2.4&968$\pm$49 \\
      19 & J120929+261331 & 0.0072 & J120917+261612 & 37 &3.3&570$\pm$22 \\
      20 & J122154+042837 & 0.0052 & J122138+043026 & 26 &0.9&2776$\pm$38 \\
      21 & J122255+154929 & 0.0053 & J122330+154507 & 39 &2.5&$<$~94 \\
      22 & J123106+120150 & 0.0260 & J123113+120307 & 83 &3.4&1191$\pm$36 \\
      23 & J130130+275232 & 0.0182 & J130128+275106 & 53 &4.3&$<$~42 \\
      24 & J130330+263235 & 0.0223 & J130345+263314 & 99 &3.2&383$\pm$61 \\
      25 & J131443+260516 & 0.0431 & J131447+260624 & 96 &2.9&803$\pm$31 \\
      26 & J135428+144030 & 0.0418 & J135426+144151 & 84 &1.8&1506$\pm$16 \\
      27 & J141512+044953 & 0.0192 & J141505+044547 & 90 &4.1&620$\pm$27 \\
      28 & J143222+095551 & 0.0046 & J143216+095520 & 6 &1.1&$-$ \\
      29 & J143245+095343 & 0.0046 & J143216+095520 & 29 &2.2&$-$ \\
      30 & J152447+042057 & 0.0374 & J152447+041920 & 107 &4.2&753$\pm$38 \\
      31 & J153313+150147 & 0.0429 & J153314+150102 & 22 &0.9&6641$\pm$292 \\
      32 & J153343+150042 & 0.0059 & J153314+150102 & 55 &2.9&648$\pm$45 \\
      33 & J155807+120411 & 0.0156 & J155821+120533 & 67 &2.9&985$\pm$40 \\
      34 & J155850+125330 & 0.0347 & J155855+125555 & 109 &3.6&585$\pm$25 \\
\hline 
\enddata
\end{deluxetable*}

\begin{deluxetable*}{ccc rccc }
% Updated on 11 July 2024
\rotate
\tablenum{4}
\tablecaption{Best fit lines for the data shown in Figure~\ref{Wexcess_MHI_SFR} in the form of y=mx+c. \label{tab:Wexcess_asur_fits}}
\tablewidth{0pt}
\tablehead{
\colhead{Abscissa (x)} & \colhead{Ordinate (y) }  & \colhead{Slope (m)} & \colhead{Intercept (c)} & \colhead{Fitting method}& \colhead{Correlation$^b$} & \colhead{Sample$^c$} }
\decimalcolnumbers
\startdata
log~M$_{HI}$                                           & log~W$_{Ly\alpha}^{excess}$   &  0.2763 $\pm$ 0.0128 &  $-$2.6965 $\pm$ 0.0170 & ASURV$^a$ & $>$ 99.8\%  & DIISC, GASS\\
% SLOPE COEFF.      :    0.2763+/-  0.0128 & INTERCEPT COEFF.   :   -2.6965+/-  0.0170 
log~(M$_{HI}$/(M$_{HI}$+M$_{\star}$)) & log~W$_{Ly\alpha}^{excess}$   &  0.3362 $\pm$ 0.0145 &  0.3031 $\pm$ 0.0213 & ASURV$^a$ & $>$ 99.9\%  & DIISC, GASS\\
% SLOPE COEFF.    :    0.3362+/-  0.0145 &INTERCEPT COEFF.   :    0.3031+/-  0.0213 
log~SFR                                                  & log~W$_{Ly\alpha}^{excess}$   &  0.1039 $\pm$ 0.0219 &  $-$0.0040 $\pm$ 0.0283 & ASURV$^a$ & $>$ 99.98\%& DIISC, GASS, Halos\\
% 7/10/2024    SLOPE COEFF.       :    0.1039+/-  0.0219    &    INTERCEPT COEFF.   :   -0.0040+/-  0.0283              
log~sSFR                                                & log~W$_{Ly\alpha}^{excess}$   &  0.1247 $\pm$ 0.0598 &   1.3187 $\pm$ 0.0621 & ASURV$^a$ & $>$ 99.99\%& DIISC, GASS, Halos\\
% 7/10/2024    SLOPE COEFF.       :    0.1247  +/-  0.0598  &       INTERCEPT COEFF.   :    1.3187 +/- 0.0621
\enddata
\tablecomments{$^a$ Linear regression fits by Schmitt's method as implemented in the survival analysis package, ASURV \citep{asurv}.\\
$^b$ Represents the confidence that a correlation exists\\
$^c$ DIISC stands for COS-DIISC, GASS for COS-GASS, and Halos for COS-Halos }
\end{deluxetable*}

\end{document}